\documentclass[fleqn,usenatbib]{mnras}

\usepackage{newtxtext,newtxmath}

\usepackage[T1]{fontenc}

\DeclareRobustCommand{\VAN}[3]{#2}
\let\VANthebibliography\thebibliography
\def\thebibliography{\DeclareRobustCommand{\VAN}[3]{##3}\VANthebibliography}

\usepackage{graphicx} 
\usepackage{amsmath} 
\usepackage{xcolor} 
\usepackage{soul} 

\usepackage{longtable}
\usepackage{caption}
\usepackage{subfig}

\newcommand{\nimg}{n_{\rm img}}
\newcommand{\Dori}{\mathcal{D}}
\newcommand{\Denh}{\mathcal{D}_{\rm shp}}
\newcommand{\Penh}{{\rm PSF}_{\rm shp}}
\newcommand{\chiflux}{\chi^2_{\rm flux}}
\newcommand{\imgsep}{{\rm sep.}}

\newcommand{\Gav}{G_{\rm av}}
\newcommand{\Nps}{N_{\rm PS}}
\newcommand{\Ngal}{N_{\rm GAL}}

\newcommand{\bd}{\begin{displaymath}}
\newcommand{\ed}{\end{displaymath}}
\newcommand{\be}{\begin{equation}}
\newcommand{\ee}{\end{equation}}
\newcommand{\beaa}{\begin{eqnarray*}}
\newcommand{\eeaa}{\end{eqnarray*}}
\newcommand{\bea}{\begin{eqnarray}}
\newcommand{\eea}{\end{eqnarray}}

\newcommand{\sref}[1]{Section~\ref{#1}}
\newcommand{\aref}[1]{Appendix~\ref{#1}}
\newcommand{\fref}[1]{Fig.~\ref{#1}}
\newcommand{\fsref}[1]{Figs.~\ref{#1}}
\newcommand{\tref}[1]{Table~\ref{#1}}
\newcommand{\tsref}[1]{Tables~\ref{#1}}
\newcommand{\eref}[1]{equation~(\ref{#1})}

\title[SuGOHI-q: Lensed quasar search in HSC DR4]{Survey of Gravitationally Lensed Objects in HSC Imaging (SuGOHI). IX. Discovery of Strongly Lensed Quasar Candidates}

\author[J.~H.~H.~Chan et al.]{
James~H.~H.~Chan,$^{1,2}$\thanks{E-mail: \url{jchan@amnh.org}}
Kenneth~C.~Wong,$^{3}$
Xuheng~Ding,$^{4}$
Dani~Chao,$^{5}$
I-Non~Chiu,$^{6}$
Anton~T.~Jaelani,$^{7,8}$
\newauthor
Issha~Kayo,$^{9}$
Anupreeta~More,$^{4,10}$
Masamune~Oguri,$^{11,12}$
and Sherry~H.~Suyu,$^{13,14,15}$ 
\\
$^{1}$Department of Astrophysics, American Museum of Natural History, Central Park West and 79th Street, NY 10024-5192, USA\\
$^{2}$Department of Physics and Astronomy, Lehman College of the CUNY, Bronx, NY 10468, USA\\
$^{3}$National Astronomical Observatory of Japan, 2-21-1 Osawa, Mitaka, Tokyo 181-8588, Japan\\
$^{4}$Kavli Institute for the Physics and Mathematics of the Universe (Kavli IPMU, WPI), The University of Tokyo, Chiba 277-8583, Japan\\
$^{5}$Ru{\dj}er Bo\v{s}kovi\'c Institute, Bijeni\v{c}ka Cesta 54, 10000, Zagreb, Croatia\\
$^{6}$Department of Astrophysics, National Cheng Kung University, No.1, University Road, Tainan City 701, Taiwan\\
$^{7}$Astronomy Research Group and Bosscha Observatory, FMIPA, Institut Teknologi Bandung, Jl. Ganesha 10, Bandung 40132, Indonesia\\
$^{8}$U-CoE AI-VLB, Institut Teknologi Bandung, Jl. Ganesha 10, Bandung 40132, Indonesia\\
$^{9}$Department of Liberal Arts, Tokyo University of Technology, Ota-ku, Tokyo 144-8650, Japan\\
$^{10}$The Inter-University Center for Astronomy and Astrophysics, Post bag 4, Ganeshkhind, Pune, 411007, India\\
$^{11}$Center for Frontier Science, Chiba University, 1-33 Yayoi-cho, Inage-ku, Chiba 263-8522, Japan\\
$^{12}$Department of Physics, Graduate School of Science, Chiba University, 1-33 Yayoi-Cho, Inage-Ku, Chiba 263-8522, Japan\\
$^{13}$Technical University of Munich, TUM School of Natural Sciences, Department of Physics,  James-Franck-Stra{\ss}e 1, 85748 Garching, Germany\\
$^{14}$Max-Planck-Institut f\"ur Astrophysik, Karl-Schwarzschild-Str. 1, 85748 Garching, Germany\\
$^{15}$Academia Sinica Institute of Astronomy and Astrophysics (ASIAA), 11F of ASMAB, No.1, Section 4, Roosevelt Road, Taipei 10617, Taiwan\\
}

\date{Accepted XXX. Received \today; in original form 10 April 2023}

\pubyear{2023}

\begin{document}
\label{firstpage}
\pagerange{\pageref{firstpage}--\pageref{lastpage}}
\maketitle

\begin{abstract}
We report the discovery of new lensed quasar candidates in the imaging data of the Hyper Suprime-Cam Subaru Strategic Program (\href{https://hsc-release.mtk.nao.ac.jp/doc/}{HSC-SSP}) DR4, covering $1\,310\deg^2$ of the sky with seeing of $\approx0.6\arcsec$. In addition to two catalogs of MILLIQUAS and AllWISEAGN, which contain confirmed and candidate quasars, we preselect quasar sources using color cuts from the HSC ($grizy$) and unWISE ($W1+W2$) photometric data based on SDSS spectroscopic catalogs. We search for the presence of multiple point sources with similar color through the convolution of the Laplacian of the preselected quasar image cutouts with the Laplacian of the point spread function, resulting in a reduction of lens candidates from 1\,652\,329 to 121\,511 (7.4\%). After visual binary classification, we grade 6\,199 (0.4\%) potential lenses on a scale of 0 to 3, with 3 indicating a lens and 0 indicating a non-lens. Finally we obtain 162 lens candidates with an average grade of $\geq2$, and among them, we successfully recover 18 known lenses. By fitting the light distribution and removing the known contaminants, we discover that 57 new systems contain at least two point sources and a galaxy in between, including 10 possible quadruply lensed quasars. This new sample of lens candidates exhibits a median separation of $1.26\arcsec$ and a magnitude limit of $i\approx22$. Spectroscopic or high-resolution imaging follow up on these newly discovered lensed quasar candidates will further allow their natures to be confirmed. 
\end{abstract}


\begin{figure*}
\centering
\includegraphics[scale=0.7]{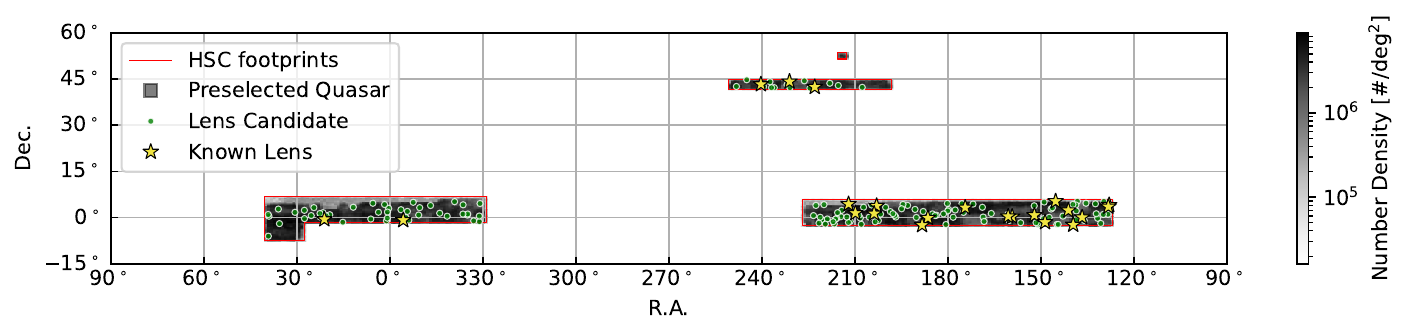}
\caption{
Population on the HSC DR4 S21A footprints. 
The colorbar represents the number density of preseleted quasar sources.
The lens candidates ($\Gav\geq2$) found in this work and the known lensed quasars are labelled as green dots and yellow stars, respectively. 
}
\label{fig:hsc}
\end{figure*}



\section{Introduction} 
\label{sec:intro}
Gravitationally lensed quasars are valuable astronomical objects that offer unique opportunities for studying galaxy evolution and cosmology. 
This phenomenon occurs when a foreground galaxy bends the light coming from a distant quasar, through the gravitational lensing effect. 
In the case of strong lensing, this can produce highly distorted, magnified, and even multiple images of the same quasar source and its host galaxy.
The image configurations of these systems provide insights into the mass structures and sub-structures of the lensing galaxies, which can further inform our understanding of galaxy evolution in the universe \citep[e.g.][]{SuyuEtal12,Dalal&Kochanek02,VegettiEtal12,NierenbergEtal17,GilmanEtal19}. 
Additionally, the time delays between the lensed images can be used to infer the Hubble constant, $H_0$, a crucial cosmological parameter that determines the size, age, and critical density of the universe \citep[e.g.][]{Refsdal64,ChenEtal19,WongEtal19}.
Given the current discrepancy between $H_0$ measurements from the early and late universe, independent measurements from lensed quasars are of vital importance \citep[e.g.][]{VerdeEtal19,DiValentinoEtal21,Freedman21}. Therefore, lensed quasars provide a powerful tool for understanding the universe's fundamental properties, making them a subject of significant interest.

To date, approximately 300 gravitationally lensed quasars have been identified.
 The largest statistical sample of radio-loud lensed quasars was provided by the Cosmic Lens All-Sky Survey \citep[CLASS;][]{MyersEtal03}, which identified flat-spectrum radio sources with multiple lensed images. 
 From optical data, the SDSS Quasar Lens Search \citep[SQLS;][]{OguriEtal06,OguriEtal08,OguriEtal12,InadaEtal08,InadaEtal10,InadaEtal12} discovered 62 lenses using morphological and color selection of spectroscopically confirmed quasars. 
 \cite{JacksonEtal12} extended the SQLS sample by utilizing better image quality from the UKIRT Infrared Deep Sky Survey \citep[UKIDSS;][]{LawrenceEtal07}.

With the increasing depth and quality of ongoing surveys, it is now possible to identify lensed quasars from imaging alone without spectroscopic preselection. 
For instance, data mining on cataloged magnitudes to identify multiple quasar sources enables the discovery of lensed quasars in various surveys, such as the Dark Energy Survey \citep{AgnelloEtal18}, the Kilo Degree Survey \citep{SpinielloEtal18,KhramtsovEtal19}, and the DESI Legacy Imaging Surveys \citep{DawesEtal22,HeEtal23}. 
To avoid contaminants, \citet{LemonEtal18,LemonEtal19} used simultaneous optical-infrared extracted colors in combination with astrometric measurements from {\it Gaia}. This method has confirmed 86 new lensed quasars \citep{LemonEtal22}, which expand the previous sample size by a factor of 1.5. 
Additionally, thanks to the exceptional spatial resolution of {\it Gaia}, the cataloged image positions have been used to identify new lenses \citep{DucourantEtal18a,DucourantEtal18b,Krone-MartinsEtal18}.

\cite{ChanEtal15} demonstrated an image-based search approach using \textsc{Chitah}\ -- an automated pixel- and mass-modeling code for finding lensed quasars.
Other image-based lens finding strategies have relied on machine learning, ring-finders, and citizen science.
These lens searches aim to find lenses, including lenses at both galaxy- and cluster-mass scales, in the Survey of Gravitationally lensed Objects in HSC Imaging \citep[SuGOHI;][]{SonnenfeldEtal18,SonnenfeldEtal19,WongEtal18,ChanEtal20,SonnenfeldEtal20,JaelaniEtal20,JaelaniEtal21,WongEtal22}.
However, this has mainly had success for lensed galaxies where the extended lensed arcs are distinguishable from other astrophysical objects. 
In the case of lensed quasars, often just two point sources are present which outshine the light from the lensing galaxy, and are outnumbered by visually similar contaminants like binary stars, or quasar-star projections. 

Here, we continue a search for lensed quasars in the HSC Data Release 4 (DR4) imaging data, using the Laplacian of an image for better point source detection.
This method has been applied in the Canada-France Imaging Survey \citep[CFIS;][]{IbataEtal17} Data Release 2 (DR2) data and successfully discovered five doubly-imaged lensed quasars \citep{ChanEtal22}.
We expect that still many lenses are yet to be discovered in the HSC survey, either due to poor point source detection or strict preselection criteria of previous search methods that discarded lensed quasars from the sample. 
The resulting sample of lensed quasars presented in this work is considered a subset of the "SuGOHI-q" lens sample.

This paper is organised as follows. 
We provide descriptions of the HSC data and the quasar selection in \sref{sec:data}. 
The automated part of our lens search method is described in \sref{sec:method}, while the visual inspection is discussed in \sref{sec:inspect}. 
We outline the light and lens modelings for the high-grade candidates in \sref{sec:model}.
We present the result and discuss the individual systems in \sref{sec:result}. 
We conclude our findings in \sref{sec:conclusion}. 
All images are oriented with North up and East left. 
Optical magnitudes quoted in this paper are in the AB system, and infrared magnitudes (specifically $W1$ and $W2$ from the $WISE$ survey) in the Vega system. 

\section{Data}
\label{sec:data}
We preselect quasar sources and look into their images in order to identify lens candidates.
In this section, we first describe the HSC imaging data (\sref{subsec:hsc}), and then the selection methods for potential quasar sources (\sref{subsec:catalog}).

\subsection{HSC imaging}
\label{subsec:hsc}

The Hyper Suprime-Cam Subaru Strategic Program \citep[HSC-SSP;][]{AiharaEtal18} is a three-layered and multi-band imaging survey with the Hyper Suprime-Cam \citep[HSC;][]{MiyazakiEtal18} on the 8.2m Subaru Telescope. 
The wide layer of the HSC-SSP consists of five broadband filters ($grizy$) with a depth of $z\approx26$ ($5\sigma$ for point sources) and a pixel scale of $0.168\arcsec$.
In this work, we take the data set from Data Release 4 (DR4), which comprises data obtained up to the semester S21A and covers $1\,310~\deg^2$ in full color, including $984~\deg^2$ to the full depth. 
The footprint of HSC DR4 S21A is illustrated in \fref{fig:hsc}.
The median seeing in the $i$-band is about $0.6\arcsec$. 
The data have been processed with the pipeline \texttt{hscPipe8.4} \citep{BoschEtal18}.
In this work, we generate cutouts of science and variance images with a size of $7.56\arcsec\times7.56\arcsec$ (45~pixels on a side), and cutouts of point spread function (PSF) with size of $5.21\arcsec\times5.21\arcsec$ (31~pixels on a side).

\begin{figure*}
\centering
\includegraphics[scale=0.6]{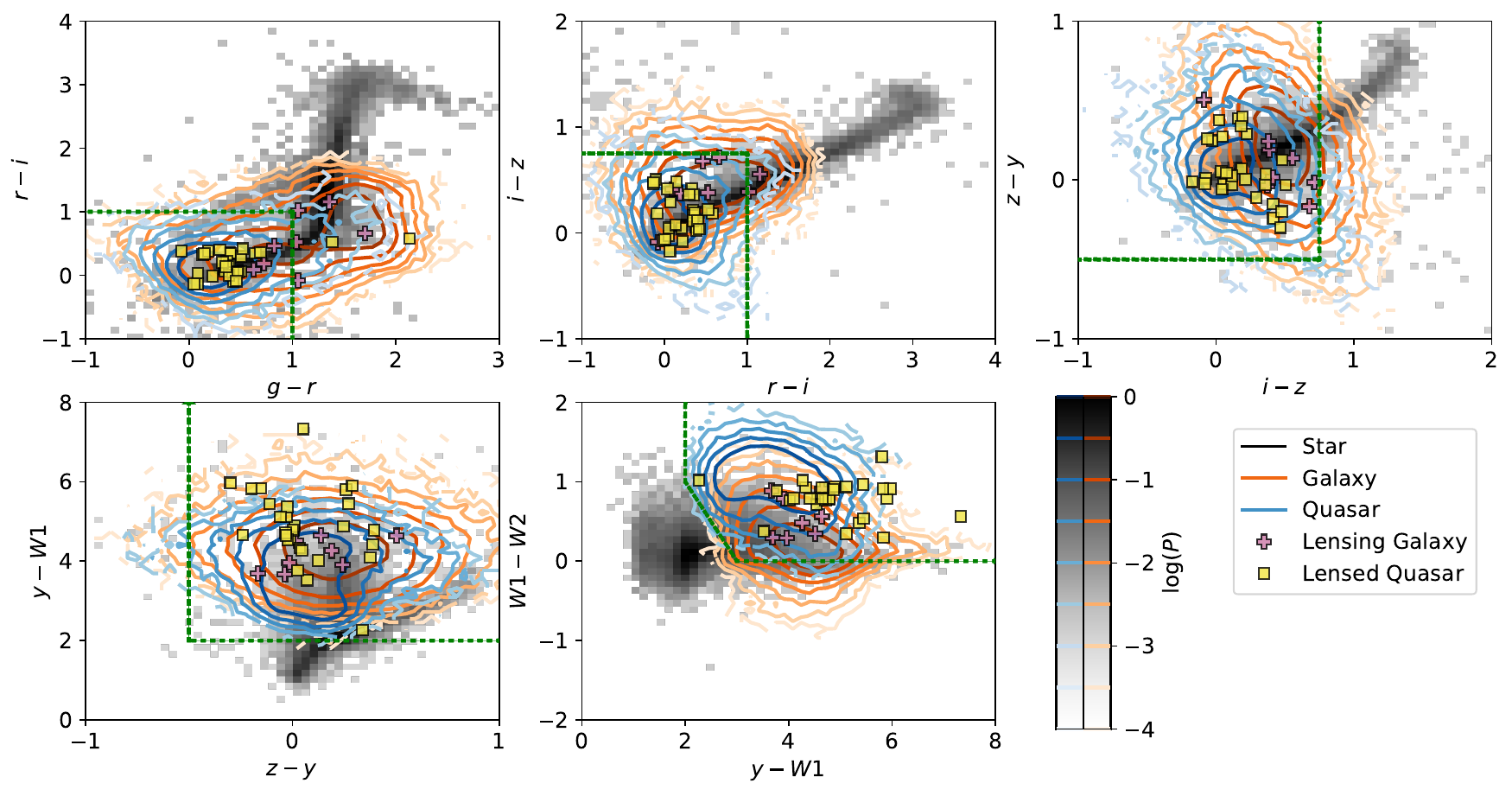}
\caption{
Color-color diagrams of the SDSS spectroscopic samples using the HSC and unWISE photometries. The color cuts applied in this work for the separation of quasars and stars are highlighted with green dotted lines, which exclude 97\% of stars and preserve 95\% of quasars.
The colorbars represent the scaled number density ($P$) with a maximum value of 1.
The known lens systems are decomposed into two components (if available in the HSC catalog): lensing galaxies and lensed quasar images, as shown in \fref{fig:known}.
Two lensed quasar images are failed within the color cuts due to high redshift (BRI0952$-$0115) and contamination by the lensing galaxy (one of the images in B1600$+$434).
}
\label{fig:sdss}
\end{figure*}

\subsection{Catalogs}
\label{subsec:catalog}

The standard way of selecting large numbers of quasar candidates is to determine “color cuts” via optical (or infrared) photometry \citep[e.g.][]{WuEtal15,PetersEtal15}. 
In order to determine the cuts, we use the spectroscopic classification in SDSS DR16 \citep{AhumadaEtal20}, containing stars, galaxies, and quasars. 
\begin{table}
  \caption{SDSS spectroscopic sample and known lenses. 
  The total number of objects before and after applying the color cuts is displayed. 
  The color cuts are detailed in \sref{subsec:catalog} and illustrated in \fref{fig:sdss}, in order to effectively separate quasars and stars.
  The sample of known lenses are decomposed into two components (if available): lensing galaxies and lensed quasar images.
  }
  \centering
  \begin{tabular}{crrr}
\hline\hline
\multicolumn{1}{c}{Type} & \multicolumn{1}{c}{$i<22$} & \multicolumn{1}{c}{color-cut} & \multicolumn{1}{c}{[\%]}\\
\hline
                Star &      13164 &        383 &       2.91 \\
              Galaxy &     254464 &      65487 &      25.74 \\
              Quasar &      60827 &      58060 &      95.45 \\
\hline
      Lensing Galaxy &          8 &          3 &      37.50 \\
       Lensed Quasar &         28 &         26 &      92.86 \\
\hline
\end{tabular}

  \label{tab:sdss}
\end{table}
We then perform a cross-match search, within radii of $1\arcsec$ and $2\arcsec$ to acquire HSC ($grizy$) photometry and unWISE ($W1/2$) photometry \citep{SchlaflyEtal19}, respectively.
A larger tolerance is applied for the infrared photometry to account for the poor seeing conditions ($>1.2\arcsec$) in the $WISE$ survey.
We only select the objects with a magnitude of $i<22$, and the color-color diagrams are shown in \fref{fig:sdss}.
For the quasar selection, we apply the color cuts of
\begin{equation}
\label{eqn:color_cuts}
\begin{array}{r@{}l}
    &g-r<1,\ \ \ r-i<1,\ \ \ i-z<0.75,\ \ \ z-y>-0.5,\\
    &y-W1>2,\ \ \ y-W2>3,\ \text{and}\ \ \ W1-W2>0,
\end{array}
\end{equation}
which enable us to remove $\approx97\%$ stars and to preserve $\approx95\%$ quasars. 
The numbers of the SDSS sample are listed in \tref{tab:sdss}.
\begin{figure}
\centering
\includegraphics[scale=0.5]{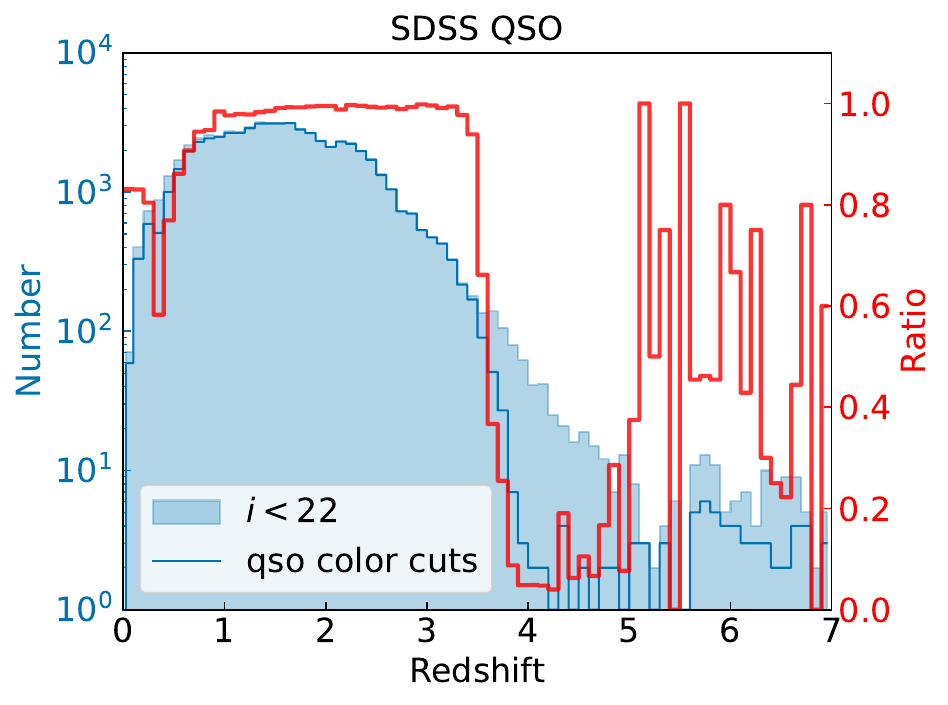}
\caption{Ratio of SDSS quasars against redshifts via the color cuts.
The sharp drop at redshift $\approx3.5$ suggests restriction on quasar sample using the color-cut selection in this work.}
\label{fig:ratio}
\end{figure}
We notice that the majority of selected quasars exhibit redshifts $z<3.5$ as shown in \fref{fig:ratio}, which is in agreement with the finding in \cite{WuEtal15}. 
The same color and magnitude cuts are applied in the HSC DR4 S21A for our quasar sample preselection.
For the reliable HSC PSF photometry, we include the conditions in the query:
\begin{equation}
\label{eqn:hsc_sql}
\begin{array}{r@{}l}
&\texttt{isprimary}\\
&\texttt{AND NOT (grizy)\_pixelflags\_interpolatedcenter}\\
&\texttt{AND NOT (grizy)\_pixelflags\_saturatedcenter}\\
&\texttt{AND NOT (grizy)\_pixelflags\_crcenter}\\
&\texttt{AND NOT (grizy)\_pixelflags\_bad}.
\end{array}    
\end{equation} 
After cross matching the unWISE catalog, we end up with 1\,587\,892 objects. 
\begin{table*}
  \caption{
  Lensed quasar selection efficiency for different catalogs.
  "Algorithm" is indicated for our classification method in \sref{sec:method}, "Phase 1" is denoted for the visual binary classification, and $\Gav$ is the average grade from the grading visual inspection.
  The percentage is relative to the "Input" number.
  The numbers include the known lensed quasars. 
  "Total" in the last row provides an assessment of the overall performance, after cross-matching the three catalogs within a $2\arcsec$ radius.
  }
  \centering
  \begin{tabular}{crrrrrrr}
\hline\hline
\multicolumn{1}{c}{Catalog} & \multicolumn{1}{c}{Input} & \multicolumn{1}{c}{Algorithm} & \multicolumn{1}{c}{[\%]}& \multicolumn{1}{c}{Phase 1} & \multicolumn{1}{c}{[\%]}& \multicolumn{1}{c}{$G_{\rm av}\geq2$} & \multicolumn{1}{c}{[\%]}\\
\hline
          HSC+unWISE &    1587892 &     121511 &       7.65 &       6111 &       0.38 &        162 &      0.010 \\
           MILLIQUAS &     150316 &       1368 &       0.91 &        376 &       0.25 &         38 &      0.025 \\
          AllWISEAGN &      49943 &        438 &       0.88 &        117 &       0.23 &         22 &      0.044 \\
\hline
               Total &    1652329 &     121880 &       7.38 &       6199 &       0.38 &        162 &      0.010 \\
\hline
\end{tabular}

  \label{tab:catalog}
\end{table*}

In addition, we include two existing quasar catalogs to slightly expand our sample size. 
The first one is the Million Quasars Catalog v7.2 \citep[hereafter MILLIQUAS;][]{Flesch21}, which provides a compilation of both confirmed and candidate quasars from the literature to 31 December 2021. 
The other catalog is the complete AllWISE Catalog of Mid-IR AGNs \citep[hereafter AllWISEAGN;][]{SecrestEtal15}, which provides the sources with their apparent magnitude distribution peak at $g\approx20$, extending to objects as faint as $g\approx26$. 

In \tref{tab:catalog}, we list the number of objects in each catalog.
To eliminate duplicates and assess the overall performance, we perform a cross-match of the three catalogs within a $2\arcsec$ radius, and the total number is displayed in the last row.

\section{Method}
\label{sec:method}

\begin{figure}
\centering
\includegraphics[scale=0.36]{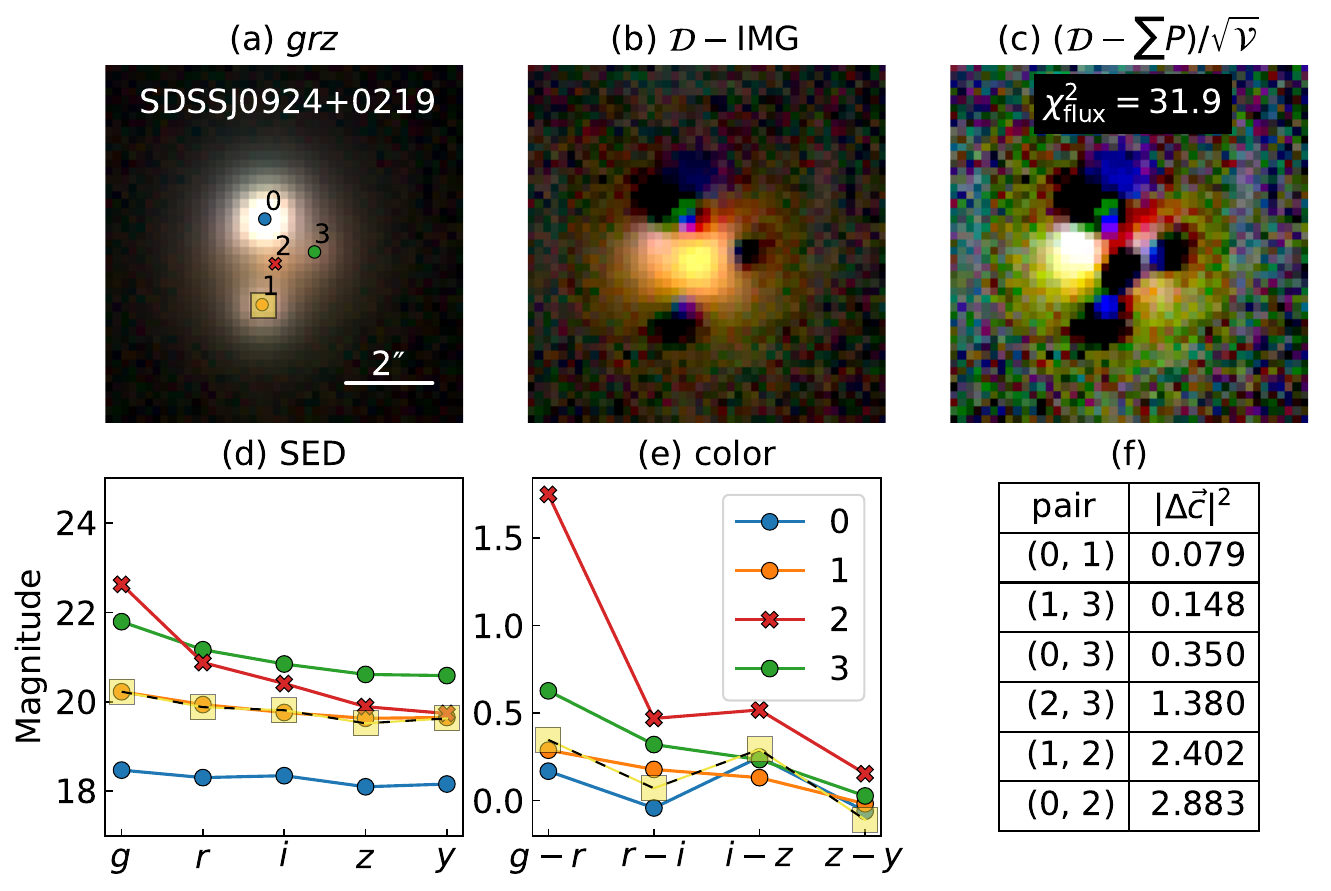}
\caption{
Output of the classification method on the known lens SDSSJ0924$+$0219: (a) composite image in the HSC \textit{grz} bands, (b) residuals upon subtraction of point sources in quasar group, (c) normalized residuals upon subtraction of the detected point sources, (d) SED of the detected point sources, (e) color of the detected point sources, (f) table of color difference.
Circles and crosses indicate the positions of quasar and non-quasar point sources, respectively.
The positions and photometry from the HSC catalog are represented in yellow squares.
}
\label{fig:chitah2}
\end{figure}

Our classification algorithm for detecting lensed quasars relies on identifying systems with multiple point sources of similar color, which is a common characteristic in a lensed quasar.
This method has been successfully applied to the Canada-France Imaging Survey (CFIS), resulting in the discovery of several new lenses \citep{ChanEtal22}.
We adopt a modified version of the procedure to better suit the HSC imaging. 
The following is a brief description of our modified procedure:
\begin{enumerate}
    \item {\it Model HSC PSF with an elliptical Moffat:}
    \label{step:psf}
    The PSFs from the HSC database can be slightly off-centered or perturbed by higher moments, which can make it difficult to accurately identify point sources (Step \ref{step:daofind}) or measure their fluxes (Step \ref{step:flux}). 
    To address this issue, we characterize the HSC database PSFs using an elliptical Moffat profile \citep{Moffat69}:
    \begin{equation}
    \label{eqn:moffat}
        I(x,y)= \frac{\beta-1}{\pi\sigma_x\sigma_y\left(1+\xi^2\right)^\beta},
    \end{equation}
    where $\sigma_x$, $\sigma_y$, and $\beta$ are seeing parameters, and $\xi$ is the scale radius with respect to the center of $(x_o,y_o)$ and the orientation $\theta$:
    \be
    \begin{aligned}
         \xi^2 = &\left[\frac{(x-x_o)\cos\theta+(y-y_o)\sin\theta}{\sigma_x}\right]^2\\
        + &\left[\frac{(x-x_o)\sin\theta-(y-y_o)\cos\theta}{\sigma_y}\right]^2.
    \end{aligned}
    \ee
    
    \item {\it Enhance image and PSF:} 
    \label{step:enhance}
    The sharpened image $\Denh$ is obtained by convolving the Laplacian of the original image $\Dori$ with the Laplacian of its corresponding PSF.
    This convolution enhances point sources in the image, expressed as:
    \be
    \begin{aligned}
    &\Denh=\Delta\Dori * \Delta\text{PSF},\\
    &\Penh=\Delta\text{PSF} * \Delta\text{PSF}, 
    \end{aligned}
    \label{eqn:enhanced}
    \ee
    where $\Delta$ denotes the Laplacian operator and $*$ is the convolution operator.
    The sharpening process reduces the Full-Width at Half-Maximum (FWHM) of the PSF by approximately 25\%, which enables the more efficient detection of point sources with small separations. (See Fig. 1 in \citealt{ChanEtal22} and Fig. 7 in \citealt{CantaleEtal16} for further details.).
    
    \item {\it Identify point sources:}
    \label{step:daofind}
    To locate the possible point sources, we use the \texttt{DAOStarFinder} python module \citep{Stetson87}.\footnote{We set the threshold to $30\ \texttt{std}$, where \texttt{std} is estimated using \texttt{sigma\_clipped\_stats} 
    The other parameters of \texttt{DAOStarFinder} are set to default values.}. 
    In addition, a 2D Gaussian fit is required in this process for the local PSF, which corresponds to $\Penh$ in \eref{eqn:enhanced}. 
    Notably, the Gaussian fit can effectively capture the characteristics of the sharpened Moffat distribution.
    The output of this process is the positions of identified point sources.
    
    \item {\it Group point sources:}
    Identifying point sources can vary in different bands. 
    To address this, we collect detections across all bands and group them within their distances less than 1.5 pixels ($0.25\arcsec$). 
    The final position is then obtained as the mean position of each group.
    
    \item {\it Remove systems with one or no identified point sources:}
    Since a single point source is unlikely to correspond to a lensed quasar system, we exclude any system with one or no point sources detected in the preceding step.
    
    \item {\it Estimate fluxes of point sources:}
    \label{step:flux}
    To determine the colors of each component, we use Powell's minimization method to estimate the flux in each band. 
    This involves placing PSFs at the positions of each component, which is obtained from Step \ref{step:daofind}. 
    The PSF is obtained using a Moffat fit, as described in Step \ref{step:psf}.
    The relevant $\chi^2$ value is calculated as follows:
    \be
    \chiflux=\frac{1}{N_{\rm band}\cdot N_{\rm pix}}\sum_{grizy}\sum_{i,j}\frac{\left[ \Dori(i,j)- \sum\limits_{k=1}^{n}P_k(i,j) \right]^2}{\mathcal{V}(i,j)},
    \ee
    where $i=1...N_x$ and $j=1...N_y$ are the pixel indices of the image cutout with dimensions of $N_x=N_y=45$, resulting in the total number of pixels $N_{\rm pix}=N_x\cdot N_y$, $N_{\rm band}=5$ is the total number of bands, $P_k(i,j)$ is the $k$-th scaled Moffat PSF at the measured position, and $\mathcal{V}(i,j)$ is the pixel variance of $\Dori(i,j)$ from the HSC database. 
    We note that the fluxes of lensing galaxies, if they are detected, are likely biased, however this in part cancels out when deriving the color of the extended component.
    
    \item {\it Remove faint point sources:}
    To avoid false detections on noise peaks, we exclude point sources that are fainter than the $5\sigma$ limiting magnitudes in all bands.
    The depths of $5\sigma$ limiting magnitudes in $grizy$ bands are estimated as 26.5, 26.1, 25.9, 25.1, and 24.4, respectively \citep{UchiyamaEtal22}.

    \item {\it Count the number of possible quasar objects:}
    We utilize the enhanced image $\Denh$ to improve the identification of point sources and to obtain more reliable PSF magnitudes than those in the HSC catalog (see \fref{fig:chitah2}).
    With the new flux measurements obtained in Step \ref{step:flux}, we apply the quasar $grizy$ color cuts as used in our preselection method (see \eref{eqn:color_cuts}) to eliminate quasars contaminated by nearby stars or galaxies. 
    When a system contains multiple potential quasar objects, we group them based on their color differences, defined as
    \begin{equation}
        {\left|\Delta\vec{c}\right|}^2 = {\left|\vec{c}_i -\vec{c}_j\right|}^2<1,
    \end{equation}
    where $\vec{c}$ is the color vector for each object, given as $\vec{c}=(g-r,r-i,i-z,z-y)$, and $i$ and $j$ are the indices of the two objects being compared.
    For simplicity, here we assign equal weight to the color difference.
    The number of possible quasar objects (dubbed as possible lensed images) in each group is counted as $\nimg$. 
    We proceed to the next step for systems with at least one group containing $2\leq\nimg\leq5$ possible lensed images. 
    The upper limit of $\nimg=5$ is empirically chosen to reduce contamination from dense star-forming galaxies or high-density stellar fields.
    We note that sometimes a quasar image can fall outside of the color cuts, for example when a quasar is contaminated by the light of its lensing galaxy or is at a high redshift, as seen in examples such as B1600$+$434 and SDSSJ1452$+$4224 in \fref{fig:known}. 
    While relaxing the quasar color cuts could prevent such cases, it would also increase the likelihood of false detections. 
    
    \item {\it Classify as a lens candidate:} 
    In order to reduce the number of false positives, we introduce three additional thresholds. 
    Firstly, we require the maximum magnitude difference between multiple point sources in each filter to be $\Delta m<2$, since large flux ratios are not compatible with lens models and not observed in known lenses \citep[see Fig. 3 in][]{ChanEtal22}.
    Secondly, lensing galaxies are typically fainter than quasar images, and even though their light can be prominent, the outskirt contribution is often insignificant after subtracting the central bulges (which we do using PSF fitting in Step \ref{step:flux}). 
    Therefore, we adopt a residual flux fit threshold of $\chiflux<200$.
    Thirdly, thanks to the high image quality of HSC, we impose an image separation ($\imgsep$) of $<3\arcsec$ and aim to expand the lens sample size in this regime.
\end{enumerate}

We calibrate this method using a sample of optically bright
known lensed quasars with HSC DR4 imaging, compiled in the Gravitationally Lensed Quasar Database \cite[GLQD; ][]{LemonEtal22}.\footnote{GLQD: \url{https://research.ast.cam.ac.uk/lensedquasars/index.html}} 
We recover 18 of 22, resulting in a recovery rate of 80\%, via color-cut preselection and the classification algorithm: $2\leq\nimg\leq5$, $\Delta m<2$, $\imgsep<3\arcsec$, and $\chiflux<200$.
The results are shown in \tref{tab:known} and \fref{fig:known}. 
We note that the measurements in the HSC catalog can be blended, leading to ambiguity in separating lensed quasar images and lensing galaxies (illustrated as yellow squares and pink crosses in \fref{fig:known}).
This test demonstrates as well our improved ability to disentangle multiple components within a lens system.
The four failures are listed as below:
\begin{enumerate}
    \item J0941$+$0518: wide image separation $\approx 6\arcsec$;
    \item B1600$+$434: slightly off quasar color in image 1, due to lensing galaxy light;
    \item BRI0952$-$0115: high redshift quasar mis-classified by the color-cut selection ($z=4.5$);
    \item SDSSJ1452$+$4224: high redshift quasar mis-classified by the color-cut selection ($z=4.8$). 
\end{enumerate}

\begin{table}
  \vskip -10pt
  \caption{
  Summary of the known lensed quasars.
  }
  \centering
  \small
  \begin{tabular}{rrrccr}
\hline\hline
\multicolumn{1}{c}{Name} & \multicolumn{1}{c}{R.A. [$\deg$]} & \multicolumn{1}{c}{Dec. [$\deg$]} & $\nimg$ & $G_{\rm av}$ & Ref. \\
\hline
        J0124$-$0033 & $   21.2394$ & $   -0.5533$ &   $2$ &   3.0 & [11] \\
      SDSSJ0832+0404 & $  128.0710$ & $    4.0679$ &   $2$ &   2.5 & [6] \\
          J0833+0331 & $  128.3369$ & $    3.5247$ &   $2$ &   2.5 & [13] \\
          J0907+0003 & $  136.7937$ & $    0.0559$ &   $2$ &   3.0 & [14] \\
        J0918$-$0220 & $  139.6806$ & $   -2.3354$ &   $2$ &   2.5 & [12] \\
      SDSSJ0924+0219 & $  141.2325$ & $    2.3236$ &   $3$ &   2.5 & [3] \\
          J0941+0518 & $  145.3439$ & $    5.3066$ &   $2$ &     - & [10] \\
          J1008+0046 & $  152.1932$ & $    0.7724$ &   $2$ &   3.0 & [13] \\
          J1037+0018 & $  159.3665$ & $    0.3057$ &   $2$ &   2.5 & [13] \\
       KIDS1042+0023 & $  160.6553$ & $    0.3839$ &   $2$ &   3.0 & [14] \\
      SDSSJ1138+0314 & $  174.5155$ & $    3.2494$ &   $4$ &   3.0 & [7] \\
    SDSSJ1226$-$0006 & $  186.5334$ & $   -0.1006$ &   $2$ &   2.5 & [7] \\
        J1233$-$0227 & $  188.4219$ & $   -2.4604$ &   $2$ &   3.0 & [13] \\
      SDSSJ1332+0347 & $  203.0942$ & $    3.7944$ &   $2$ &   2.0 & [5] \\
      SDSSJ1335+0118 & $  203.8950$ & $    1.3015$ &   $2$ &   2.5 & [4] \\
          J1359+0128 & $  209.9342$ & $    1.4693$ &   $2$ &   3.0 & [12] \\
          J1408+0422 & $  212.1406$ & $    4.3747$ &   $2$ &   3.0 & [13] \\
      SDSSJ1524+4409 & $  231.1901$ & $   44.1638$ &   $2$ &   3.0 & [6] \\
           B1600+434 & $  240.4185$ & $   43.2799$ &   $1$ &     - & [2] \\
    ULASJ2343$-$0050 & $  355.7998$ & $   -0.8429$ &   $2$ &   2.0 & [8] \\
      BRI0952$-$0115 & $  148.7504$ & $   -1.5019$ &     - &     - & [1] \\
      SDSSJ1452+4224 & $  223.0479$ & $   42.4082$ &     - &     - & [9] \\
\hline
\end{tabular}
[1]~\cite{McMahonEtal92}, [2]~\cite{JacksonEtal95}, [3]~\cite{InadaEtal03}, [4]~\cite{OguriEtal04}, [5]~\cite{MorokumaEtal07}, [6]~\cite{OguriEtal08}, [7]~\cite{InadaEtal08}, [8]~\cite{JacksonEtal08}, [9]~\cite{MoreEtal16}, [10]~\cite{WilliamsEtal18}, [11]~\cite{LemonEtal19}, [12]~\cite{JaelaniEtal21}, [13]~\cite{LemonEtal22}, [14]~\href{https://research.ast.cam.ac.uk/lensedquasars/index.html}{GLQD}
  \label{tab:known}
\end{table}
\begin{figure}
\centering
\includegraphics[scale=0.31]{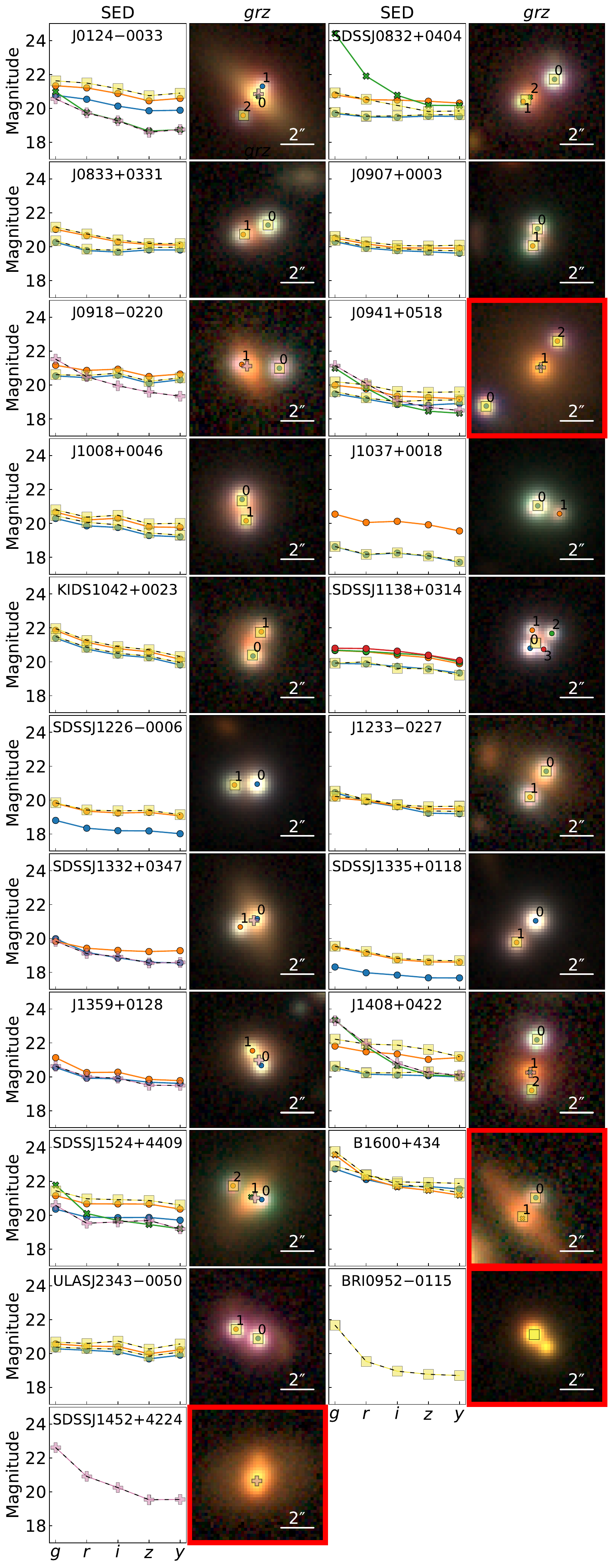}
\caption{
Known lens test for the classification method. 
The labels are same as \fref{fig:chitah2} (a) and (d). 
The measurements from the HSC catalog are denoted in yellow squares and pink crosses, representing two components of a lens system: lensed quasar image and lensing galaxy, respectively.
The lenses which are not able to be recovered are highlighted with the red frames. 
BRI0952$-$0115 and SDSSJ1452$+$4224 fail the color-cut preselection due to high redshift quasars.
J0941$+$0518 has a wide image separation $\approx6\arcsec$, while B1600$+$434 exhibits an image with slightly off quasar color due to lens light contamination.
}
\label{fig:known}
\end{figure}

After performing calibration on the method, we then apply to the preselected quasar candidates from \sref{sec:data}, resulting in a reduced sample size of 121\,511 systems (7.4\%) for visual inspection. 
This represents a significant reduction from the original catalog size. 
The computational time for analyzing each system is approximately 2 seconds.

\section{Visual Inspection}
\label{sec:inspect}

\begin{figure}
\centering
\includegraphics[scale=0.31]{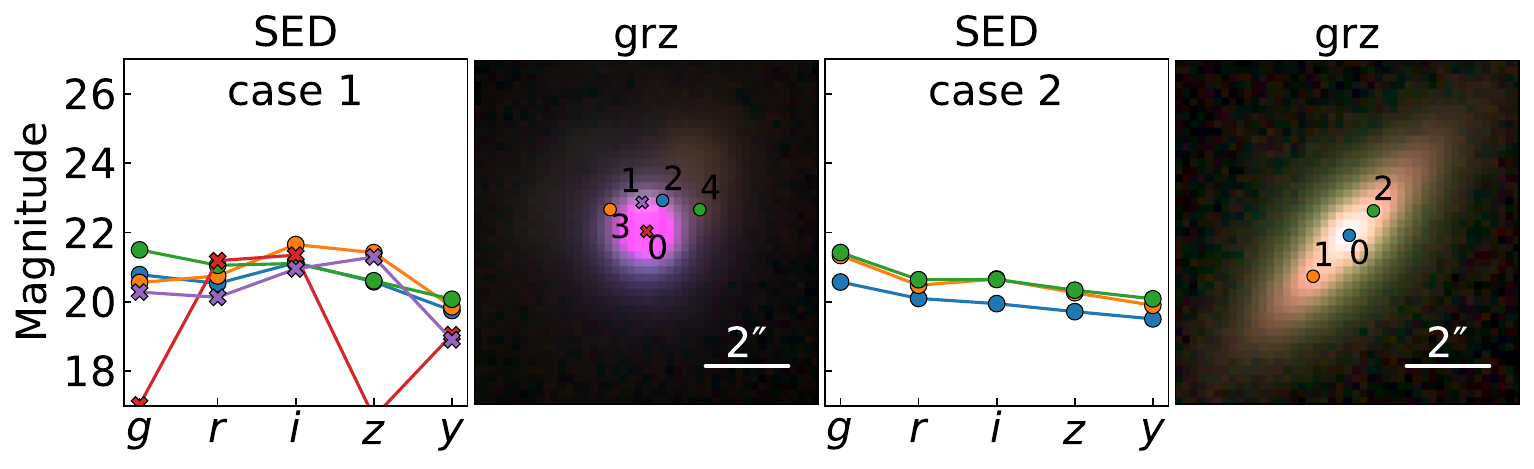}
\caption{
Examples of objects classified as "other" during the visual binary classification, due to bad pixels (case~1), and extended galaxies (case~2).
}
\label{fig:false}
\end{figure}

To handle the large number of lens candidates (121\,511), we propose a two-phase visual inspection process. 
In this process, we intentionally include quasar pairs as candidates because they are a byproduct of this work and can provide valuable insights into the existence of supermassive black hole pairs and galaxy mergers in the early universe \citep[e.g.][]{GouldingEtal19,Casey-ClydeEtal21}.
During the first phase of visual inspection, we perform a binary classification of "possible lens or quasar pair" and "other", which is carried out by the first three authors. 
JHHC inspects the whole sample, while KCW and XD share it equally, ensuring that each system is reviewed by two independent inspectors.
At this stage we take a conservative approach, focusing on maintaining possible lenses or quasar pairs.
Nevertheless, we ensure to remove any apparent non-target objects that could lead to false positives.
Each system's image for inspection is generated from the results of the classification algorithm, as shown in \fref{fig:chitah2}. 
To better identify faint, compact lensing galaxies, we take residual images obtained by subtracting the possible lensed images.
The majority of candidates are classified as "other" in this phase due to one or several of the following reasons: 
\begin{itemize} 
    \item[--] saturated pixels in the cutout, as they can produce false point source detections in our algorithm, especially in the existing catalogs;
    \item[--] obvious extended galaxies or non-point sources in the images, such as the host galaxy of a low-redshift AGN or a galaxy merger.
\end{itemize}
Each case is illustrated in \fref{fig:false}.
By taking the union of the three sets from the first three inspectors, we identify 6\,199 candidates for the next phase.

The second phase involves a quaternary classification in which each inspector grades candidates from 0 to 3 using only integer values, according to the following criteria:
\begin{itemize}
    \item[3:] very likely to be a lens, requiring resolved spectroscopy for confirmation;
    \item[2:] blue point sources at modest separation with similar quasar-like colors and without an obvious lensing galaxy;
    \item[1:] same as the previous grade but with slightly high values of PMSIG and AEN (definitions explained later in this section), as well as larger differences in colors;
    \item[0:] a system that likely contains a star, star-forming galaxy, or a very low-redshift quasar.
\end{itemize}
To avoid bias, we assign each system two grades. First, JHHC goes through the entire sample and divides it into nine batches, which are shared by JHHC, KCW, DC, IC, ATJ, IK, AM, MO, and SHS. 
JHHC grades some candidates twice to check grade consistency. 
For each candidate, we inspect any relevant data of the system, including SDSS images and spectra, DECaLS images,\footnote{DECaLS: \url{https://www.legacysurvey.org}} \citep{DeyEtal19} unWISE images \citep{MeisnerEtal18}, catalogs of VizieR \citep{OchsenbeinEtal00}, NED,\footnote{The NASA/IPAC Extragalactic Database (NED) is operated by the Jet Propulsion Laboratory, California Institute of Technology, under contract with the National Aeronautics and Space Administration.} and Simbad \citep{WengerEtal00}. 
We also consider {\it Gaia} DR3 proper motions \citep[converted to the PMSIG parameter defined in][]{LemonEtal19}, and astrometric excess noise \citep[AEN;][]{gaia,gaiadr3}.
A large PMSIG value ($\gtrsim 10~\sigma$) strongly suggests that the system contains a star, while a high AEN value ($\gtrsim 10$ mas) may indicate a candidate star-forming galaxy \citep[see Figs. 1 and 2 in][]{LemonEtal19}.
We acknowledge that the grading process is subjective and time-consuming, and we note that only JHHC has seen images of confirmed lensed quasars. 
While the two grades assigned to each system are usually nearly identical, there are cases where there is a grade difference ($\Delta G \geq 2$), likely due to personal grading experience or human fatigue. 
After discussing these candidates, the percentage of cases with $\Delta G \geq 2$ drops from 7\% to <3\%. 
In the end, we take the average of the two grades assigned to each system ($\Gav$), resulting in 31 systems with $\Gav=3$, 42 systems with $\Gav=2.5$, and 89 systems with $\Gav=2$.
We then perform detailed modeling for a total of 162 candidates in \sref{sec:model}.

\section{Light and lens modeling analysis}
\label{sec:model}

A lensed quasar system is expected to have a lensing galaxy that is surrounded by multiple lensed images of the quasar. 
However, in most cases, the lensing galaxy is much fainter than the quasar and can hardly be seen in ground-based imaging. 
Although our classification method is effective at highlighting lensing galaxies by removing lensed quasar images (as demonstrated in \fref{fig:chitah2}), double point sources can sometimes remain, creating a residual that mimics a lensing galaxy. 
To tackle this issue, we aim to optimally subtract any discernible point sources (or single point sources and galaxies), to enhance our ability to identify a lensing galaxy or counter image in the deep, high-resolution HSC imaging.
This approach is applied to the most promising 162 systems resulting from our visual inspection.

We model point sources using a Moffat profile (from Step \ref{step:psf} in \sref{sec:method}) and galaxies using a S\'ersic profile \citep{Sersic63}. 
We manually determine the numbers of point sources ($\Nps$) and galaxies ($\Ngal$) for each system, and simultaneously fit the $grizy$ bands. 
The resulting $\chi^2$ is defined as
\begin{equation}
\chi_{\rm light}^2=\frac{1}{N_{\rm band}\cdot N_{\rm pix}}\sum_{grizy}\sum_{i,j} \frac{\left[\Dori(i,j)-\mathcal{M}(i,j)\right]^2}{\mathcal{V}(i,j)}. 
\end{equation}
The model image $\mathcal{M}$ is the sum of all components in each band:
\begin{equation}
\mathcal{M} = \sum^{\Nps}_k P_k(\vec{r}_{p,k}) + \sum^{\Ngal}_k S_k(A_k,R_{e,k},e_k,\phi_k,n_k,\vec{r}_{l,k}),
\end{equation}
where $k$ is the index of each component, $P$ is the scaled Moffat PSF at the position parameter $\vec{r}_p$, and $S$ is a scaled S\'ersic profile with the parameters of the intensity ($A$), the effective radius ($R_e$), the ellipticity ($e$), the orientation ($\phi$), the S\'ersic index ($n$), and the center ($\vec{r}_l$).
The results of the light modeling can be found at the end of the paper (see \fsref{fig:cand_A}, \ref{fig:cand_B}, and \ref{fig:cand_C}).
The measured astrometry and photometry are listed in \tref{tab:model}.

\begin{figure*}
\centering
\includegraphics[scale=0.55]{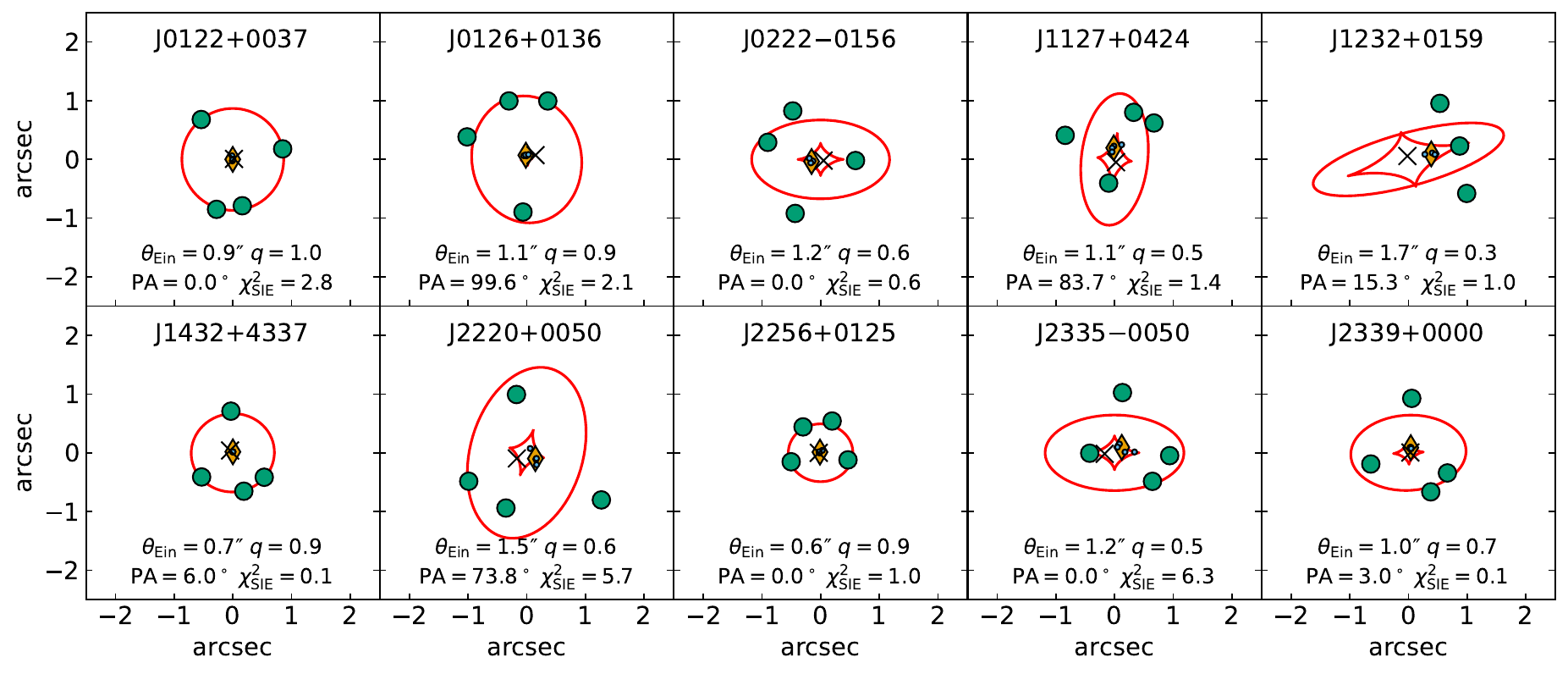}
\caption{
Lens modeling for quadruply lensed quasar candidates. 
A set of model parameters are denoted at the bottom of each panel, and the resulting critical curves and caustics are illustrated in red lines.
The galaxy position is obtained from light modeling, represented as black cross.
The lensed image positions obtained from light modeling are shown as green circles, while the mapped sources are represented by blue dots.
The modeled source position is labeled as orange diamonds.
}
\label{fig:lens_model}
\end{figure*}

We consider a few candidates with the potential to contain quadruply lensed images, and model their image configuration using a singular isothermal elliptical (SIE) lens mass distribution. 
This involves using parameters such as the Einstein radius ($\theta_{\rm Ein}$), axis ratio ($q$), position angle (PA), and lens center ($\vec{r}_{\rm SIE}$).
The resulting $\chi^2$ is calculated by comparing the modeled and observed source positions, as well as by comparing the lens mass to light centers \citep{Oguri10,ChanEtal15,ChanEtal20}, which can be described as 
\begin{equation}
 \chi_{\rm SIE}^2=\sum_k \frac{\mu_k\left|\vec{\beta}_k-\vec\beta_{\rm SIE}\right|^2}{{\sigma_k}^2}+\frac{\left|\vec{r}_{\rm SIE}-\vec{r}_{l}\right|^2}{{\sigma_l}^2}, 
\end{equation}
where $\vec{\beta}_k$ is the respective source position mapped from the
position of lensed image $k$ with the magnification $\mu_k$, and $\vec{\beta}_{\rm SIE}$ is the modeled source position evaluated as a weighted mean of $\vec{\beta}_k$,
\begin{equation}
\vec{\beta}_{\rm SIE}=\frac{\sum\limits_{k} \sqrt{\mu_k}\cdot \vec{\beta}_k}{\sum\limits_{k} \sqrt{\mu_k}}.
\end{equation}
The light center $\vec{r}_l$ is obtained from the light modeling, with the uncertainties of image and lens positions conservatively to be the pixel scale of HSC, i.e. $\sigma_k=\sigma_l=1~{\rm pix} = 0.168\arcsec$.
Although this choice does not affect the fit, we adopt it for consistency.
The lens modeling results are illustrated in \fref{fig:lens_model}.

\section{Result}
\label{sec:result}

According to modeling results on 162 candidates with $\Gav\geq2$, we provide the final grades as 
\begin{itemize}
    \item[A:] requiring a fit with at least two point sources and one galaxy;
    \item[B:] not satisfying A but with $\Gav>2$;
    \item[C:] not satisfying A but with $\Gav=2$;
    \item[G:] likely to be a lensed galaxy. 
\end{itemize}
The results are summarized in \tsref{tab:candA}, \ref{tab:candB}, and \ref{tab:candC}, which include the maximum distance between point sources as a tentative measure of image separation.
We should note that in some cases, determining whether a source is best modeled as a point source or/and a S\'ersic profile can be ambiguous, particularly for compact sources. 
There are ten candidates identified as potentially containing quadruply lensed images (quads), with six showing promising results and four being inconclusive.
A specific classification, referred as grade G, is described in \aref{sec:grade_G}, indicating the likely presence of arcs after single galaxy fit.
We also find candidates with two widely separated point sources without a galaxy in between, which are not lenses but may be quasar pairs or nearly identical quasars. 
These have potential applications in black hole science. 
The number of such candidates with $\Gav<2$ is 774, which are not presented in this work but may be made available upon request from the authors.

\begin{figure}
\centering
\includegraphics[scale=0.4]{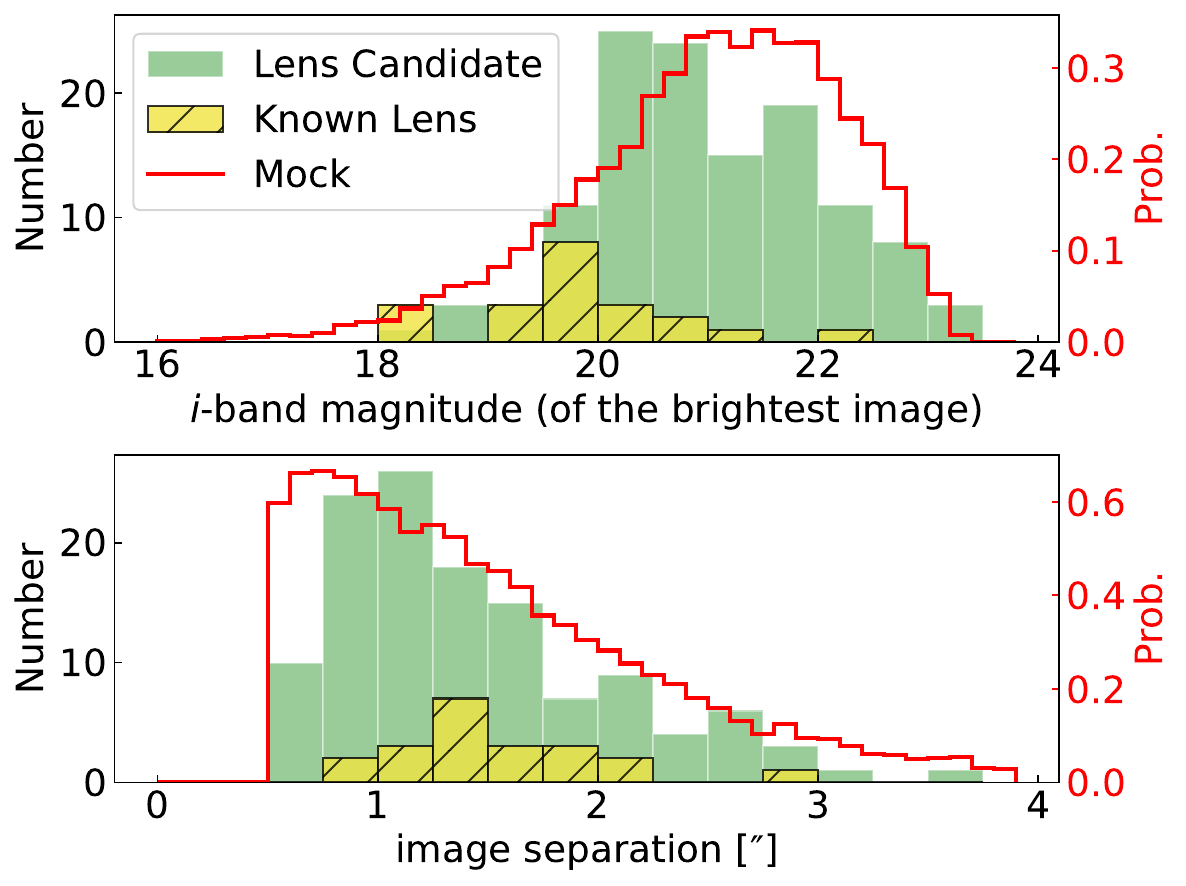}
\caption{Histograms of the $i$-band magnitude (top) and of the image separation (bottom). 
The $i$-band magnitude and image separation are obtained from the brightest putative lensed images and the maximum distance between them, respectively.
The photometry and astrometry of lens candidates (green) are measured from light modeling and those of known lenses (yellow) are from the classification algorithm.
The probability distributions of our lens candidates are consistent with those of the mocks (red) in \citet{Oguri&Marshall10}.
}
\label{fig:hist}
\end{figure}

Only 60\% of the quasars in the existing catalogs are identified in the catalog using color cuts.
However, according to our findings, the color-cut preselection is comprehensive as it includes all candidates with $\Gav\geq2$.
While existing quasar catalogs can lead to higher classification efficiency (see \tref{tab:catalog}), our results demonstrate the effectiveness of photometric selection for lensed quasar searches, especially when considering \textit{WISE} colors.
The high-resolution and deep imaging capabilities of HSC enable us to expand the sample size of low-brightness and small-image-separation lenses. 
Although we have yet to confirm the nature of our candidates, we present histograms of the lens candidate sample as a function of the $i$-band magnitude of the brightest image and the putative image separation in \fref{fig:hist}.
These probability trends closely follow those seen in the mock catalog simulations \citep{Oguri&Marshall10}, providing further support for the validity of our method.

We provide individual notes for several interesting systems, such as potential quads, grade A candidates with spectroscopic redshifts, and those with identified contaminants, although the list is not comprehensive.

\subsection{J0122$+$0037 (Quad)}
This system is a promising lens, though the lensed images do not appear as point sources. 
We conducted a tentative fit using four point sources and one galaxy and find that the image configuration can be well modeled with an SIE. 
However, an SIE with $q=1$ is less likely to produce a quad system, possibly due to the poor representation of the lensed arc feature using four point sources.
Further confirmation of its nature will require high-resolution imaging or spectroscopy.

\subsection{J0126$+$0136 (Inconclusive)}
This system could be a lensed galaxy. 
We tentatively fit with four point sources and one galaxy, but the lensing feature is not clearly pronounced.
Therefore, we note it as inconclusive.

\subsection{J0222$-$0156 (Inconclusive)}
We performed a tentative fit for this system using four point sources and one galaxy. 
Although the lens model can fit the data, we observed that the lensing feature of components A, B, and C could also be formed by two galaxies, as shown in the residual. 
Additionally, the galaxy fit at the center is not consistent with the light modeling, indicating that the results are not promising. 
Therefore, we consider this system to be inconclusive.

\subsection{J1127$+$0424 (Quad)}
This system exhibits a clear fold configuration, with two adjacent images close to the critical curve, and can be well modeled by an SIE.
However, further confirmation of its nature will require high-resolution imaging or spectral data.

\subsection{J1232$+$0159 (Quad)}
This system contains a triplet of lensed images with the same color, although the fourth image is difficult to detect. 
We model the three lensed images and find that the lens model is highly elliptical, likely due to the nearby galaxy group (see \fref{fig:cand_A}). 
High-resolution imaging or spectroscopic observations are needed to confirm the nature of this system.

\subsection{J1432$+$4337 (Quad)}
This system is a promising lens, although the lensed images do not appear to be point sources, and there is a possibility that they may be blended with lensed arcs.

\subsection{J2220$+$0050 (Inconclusive)}
This system has a galaxy spectrum with a redshift of $z=0.616$ obtained from VVDS.\footnote{VVDS: \url{https://cesam.lam.fr/vvds/}}
However, the image D appears slightly reddened and out of position, leading to a larger $\chi^2_{\rm SIE}$. 
Therefore, we classify this system as inconclusive.

\subsection{J2256$+$0125 (Inconclusive)}
Thanks to the HSC image quality, we can see a small arc in this system. 
We tentatively fit with four point sources and one galaxy, though image D is hardly detected. 
Due to its small image separation around $1\arcsec$, we note it as inconclusive.

\subsection{J2335$-$0050 (Quad)}
This system is identified as a type 2 Seyfert galaxy at redshift $z=0.438$ based on the SDSS spectrum. 
However, the fourth image (image D) is unclear and not visible in the HSC images. 
If it turns out to be a lens, we note that there is a strong flux anomaly between images A and B.

\subsection{J2339$+$0000 (Quad)}
This system has clear lensing features, including a fold configuration and visible lensed arcs in the residuals.
High-resolution imaging and/or a spectrum are needed to confirm the nature of the source

\subsection{J1338$+$0405 (Inconclusive)}
This system appears to be a doubly lensed quasar, but the identification of image C as a star with a large ${\rm PMSIG} = 36.7~\sigma$ creates uncertainty.
Therefore, we note this system as inconclusive.

\subsection{J0141$+$0039 (Double)}
This system has been confirmed as a quasar source with redshift $z=2.198$ in SDSS, and it has been identified as a strong candidate for a gravitational lens in SQLS.
Although follow-up spectroscopic observations have been conducted, its status as a lensed quasar remains uncertain due to its small separation of only about $1\arcsec$.

\subsection{J0954$-$0022 (Double)}
This system has been confirmed as a quasar source with a redshift of $z=2.267$ using the 2dF instrument on the Anglo-Australian Telescope (AAT). 
Based on our light modeling, we estimate the image separation to be $1.26\arcsec$. 
However, the lensing galaxy is obscured, making it difficult to study the lensing properties in detail.

\subsection{J1436$-$0026 (Inconclusive)}
This system is confirmed as a quasar at redshift $z=0.5615$ in the 2SLAQ catalog \citep{CroomEtal09}.
However the galaxy fit is unclear, as it appears to be brighter than the other two point sources. 
Therefore, we note it as inconclusive until further observations can provide more information.

\subsection{J1453$+$0039 (Inconclusive)}
This system is confirmed as a quasar at redshift $z=1.094$ from SDSS.
However, the galaxy fit appears to be slightly displaced from the expected position, which may be due to the presence of a nearby galaxy.
As a result, we note it as inconclusive.

\subsection{J0847$-$0013 (Contam.)}
This system was first identified as a lens candidate \citep{InadaEtal08}, but has been confirmed as a dual quasar system \citep{SilvermanEtal20}, as a result from different Mg \textsc{ii} line profiles in each component. 
Even though it requires a galaxy in light modeling, the galaxy turns out to be a host galaxy with a stellar mass $8.5\times10^{10}~M_\odot$.

\subsection{J0848$+$0115 (Contam.)}
This system was initially identified as a lens candidate \citet{InadaEtal12} with a quasar at redshift $z=0.646$. But it turns out to be a quasar+galaxy+star system (commented in GLQD).

\subsection{J0947$+$0247 (Contam.)}
This system was first identified as a lens candidate in SQLS \citep{InadaEtal10}, but later confirmed as a quasar ($z=0.643$) + star \citep{LemonEtal22}.

\subsection{J1008$+$0351 (Contam.)}
This system was first identified as a lens candidate in SQLS \citep{InadaEtal08}, but later confirmed as a quasar pair with redshifts of $z=1.745$ and $z=1.740$ \citep{KayoOguri12}.

\subsection{J1406$-$0112 (Contam.)}
This system with a quasar at redshift $z=1.154$ was initially identified as a lens candidate in SQLS \citep{InadaEtal08}.
However, it was later classified as a contaminant in GLQD due to the absence of a lensing galaxy.

\subsection{J2205$+$0031 (Contam.)}
This system was classified as a lens candidate \citep{TreuEtal18}, but later commented as a quasar pair with redshifts of $z=1.653$ and $z=1.869$ in GLQD.

\section{Summary}
\label{sec:conclusion}
In this work, we introduce new candidates for lensed quasars in HSC DR4 S21A, covering an area of approximately $1300~\deg^2$. 
We draw the following conclusions:
\begin{itemize}
    \item To identify possible quasar sources, we utilize HSC+unWISE photometry ($grizy+W1+W2$) and apply color cuts that enable us to exclude the vast majority of stars while including most quasars with redshifts $z < 3.5$.
    In addition, we incorporate two existing quasar catalogs (MILLIQUAS and AllWISEAGN), resulting in a total of 1\,652\,329 objects.
    \item Using an efficient classification method developed in \cite{ChanEtal22}, modified to better suit the HSC imaging, we are able to recover 18 of 22 known lenses.
    \item We visually inspected 121\,511 candidates and identified 6\,199 systems with lensing features, which are graded on a scale of 0 (non-lens) to 3 (lens).
    \item For the 162 candidates with an average grade of $\Gav\geq2$, we conducted light modeling, and 10 systems showing quadruply lensed images were further modeled to determine the image configuration.
    \item Based on the light and lens modeling results, we assigned final grades of A, B, and C, resulting in 57, 15, and 53 candidates, respectively, after the known contaminants are removed. 
\end{itemize}
Despite the lack of spectroscopic information to confirm the nature of our candidates, a viable strategy for follow-up observations is through long-slit spectroscopy using existing instruments.
Additionally, these candidates present excellent targets for the Prime Focus Spectrograph (PFS), which covers the same area as the HSC and offers the potential to obtain high-quality spectroscopic data.
By analyzing these data, we can gain a more accurate understanding of the properties of the lensing systems identified in this study.
Our successful identification of promising lensed quasar candidates in this work indicates that our classification method can be applied to higher resolution imaging data in the future, such as Euclid and the Vera C. Rubin Observatory's Legacy Survey of Space and Time (LSST), and could lead to the discovery of lenses with the narrowest separation on the sky.

\section*{Acknowledgements}

J.~H.~H.~C. thanks F.~Courbin, C.~Lemon, Y.-P.~Shu, J.~Silverman, Q.~Minor, and C.~Rusu for useful discussions.
This research was made possible by the generosity of Eric and Wendy Schmidt by recommendation of the Schmidt Futures program.
I.~C. is supported by the National Science and Technology Council in Taiwan (Grant NSTC 111-2112-M-006-037-MY3).
A.~T.~J. is supported by the Program Riset ITB 2023.
S.~H.~S. thanks the Max Planck Society for support through the Max Planck Research Group and the Max Planck Fellowship.
This work was supported by JSPS KAKENHI Grants: Number JP20K14511 for K.~C.~W., Number JP20K04016 for I.~K., and Numbers JP22H01260, JP20H05856, and JP20H00181 for M.~O.

The Hyper Suprime-Cam (HSC) collaboration includes the astronomical communities of Japan and Taiwan, and Princeton University. The HSC instrumentation and software were developed by the National Astronomical Observatory of Japan (NAOJ), the Kavli Institute for the Physics and Mathematics of the Universe (Kavli IPMU), the University of Tokyo, the High Energy Accelerator Research Organization (KEK), the Academia Sinica Institute for Astronomy and Astrophysics in Taiwan (ASIAA), and Princeton University. Funding was contributed by the FIRST program from the Japanese Cabinet Office, the Ministry of Education, Culture, Sports, Science and Technology (MEXT), the Japan Society for the Promotion of Science (JSPS), Japan Science and Technology Agency (JST), the Toray Science Foundation, NAOJ, Kavli IPMU, KEK, ASIAA, and Princeton University.
 
This paper is based on data collected at the Subaru Telescope and retrieved from the HSC data archive system, which is operated by Subaru Telescope and Astronomy Data Center (ADC) at NAOJ. Data analysis was in part carried out with the cooperation of Center for Computational Astrophysics (CfCA) at NAOJ. We are honored and grateful for the opportunity of observing the Universe from Maunakea, which has the cultural, historical and natural significance in Hawaii.
 
This paper makes use of software developed for Vera C. Rubin Observatory. We thank the Rubin Observatory for making their code available as free software at \url{http://pipelines.lsst.io/}. 
 
The Pan-STARRS1 Surveys (PS1) and the PS1 public science archive have been made possible through contributions by the Institute for Astronomy, the University of Hawaii, the Pan-STARRS Project Office, the Max Planck Society and its participating institutes, the Max Planck Institute for Astronomy, Heidelberg, and the Max Planck Institute for Extraterrestrial Physics, Garching, The Johns Hopkins University, Durham University, the University of Edinburgh, the Queen’s University Belfast, the Harvard-Smithsonian Center for Astrophysics, the Las Cumbres Observatory Global Telescope Network Incorporated, the National Central University of Taiwan, the Space Telescope Science Institute, the National Aeronautics and Space Administration under grant No. NNX08AR22G issued through the Planetary Science Division of the NASA Science Mission Directorate, the National Science Foundation grant No. AST-1238877, the University of Maryland, Eotvos Lorand University (ELTE), the Los Alamos National Laboratory, and the Gordon and Betty Moore Foundation.

\section*{Data Availability}
 
The data underlying this article will be shared on reasonable request to the corresponding author.



\bibliographystyle{mnras}
\bibliography{reference} 

\begin{appendix}

\section{Grade G candidates}
\label{sec:grade_G}

We identify several lensed galaxy candidates, including J0004$-$0103, which has been previously rediscovered in multiple studies. 
Although our classification algorithm is primarily designed for lensed quasars, it can also identify these lensed galaxies. 
The results of lensed galaxy candidates are presented in \fref{fig:candG} and \tref{tab:candG}.

\section{Result from light modeling}
\label{sec:photometry}

For lens candidates with $\Gav\geq2$ listed in \tref{tab:model}, we measure the HSC astrometry and photometry via light modeling (see \sref{sec:model}).

\end{appendix}
\clearpage
\setcounter{figure}{8}   
\setcounter{table}{3}   
\begin{figure*}
\includegraphics[scale=0.24]{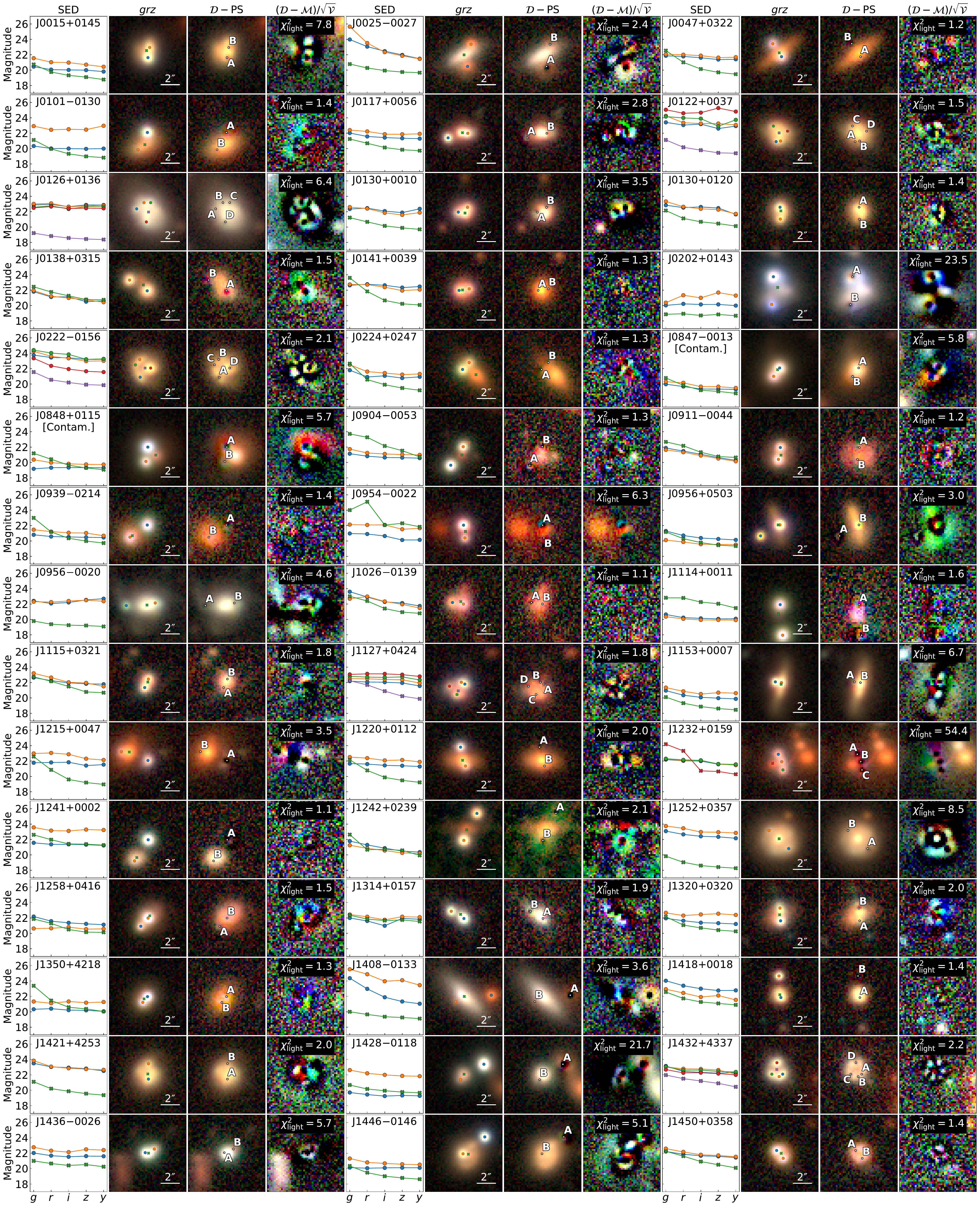}
\caption{
Lensed candidates classified as grade A. 
Each candidate contains "SED", "\textit{grz}", "$\mathcal{D}-\text{PS}$", and "$(\mathcal{D}-\mathcal{M})/\sqrt{\mathcal{V}})$" from light modeling. 
The labels in "SED" and "\textit{grz}" are the same as \fref{fig:chitah2} (a) and (d).
The results of light modeling for each component are listed in \tref{tab:model}.
The contaminants are noted under their names.
}
\label{fig:cand_A}
\end{figure*}
\begin{figure*}
\ContinuedFloat
\includegraphics[scale=0.24]{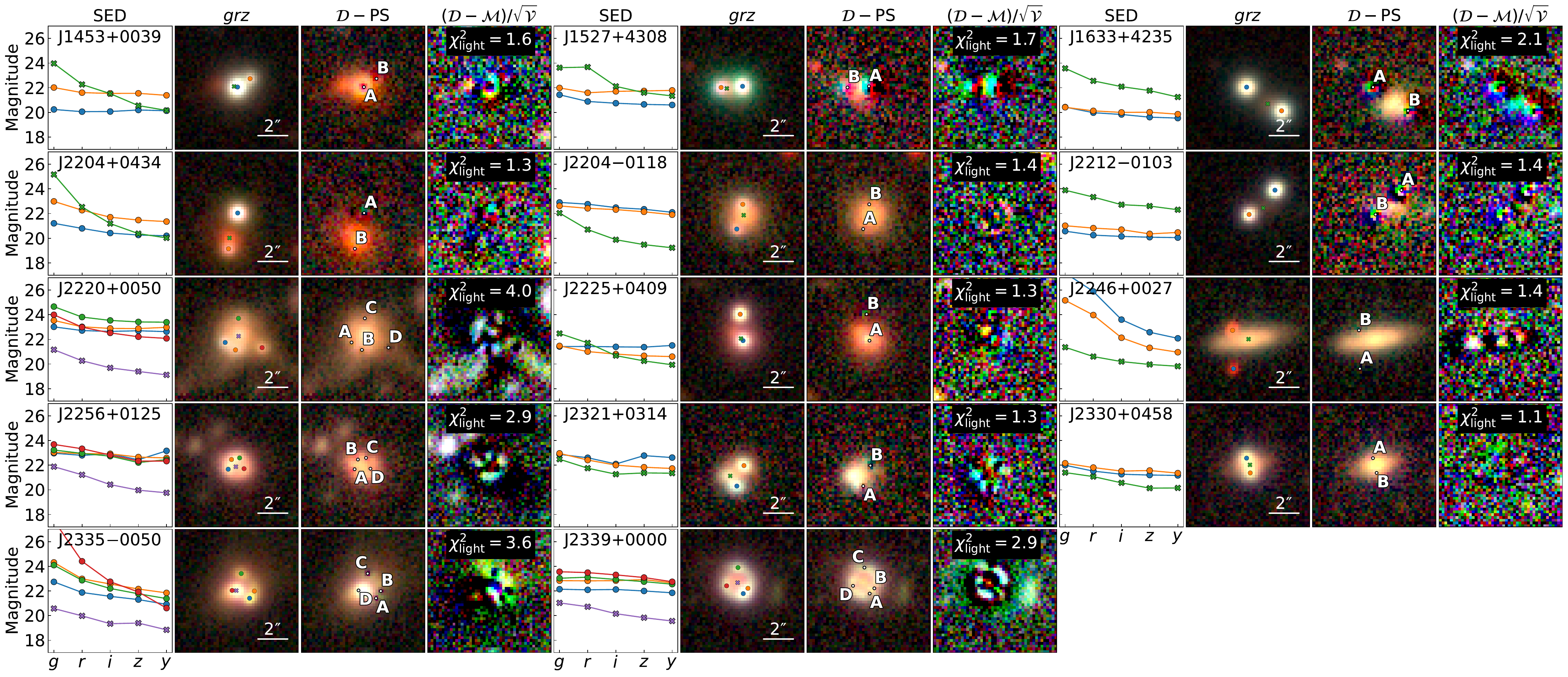}
\vskip -10pt
\caption{\textit{Continued.}
}
\end{figure*}
\begin{figure*}
\centering
\includegraphics[scale=0.24]{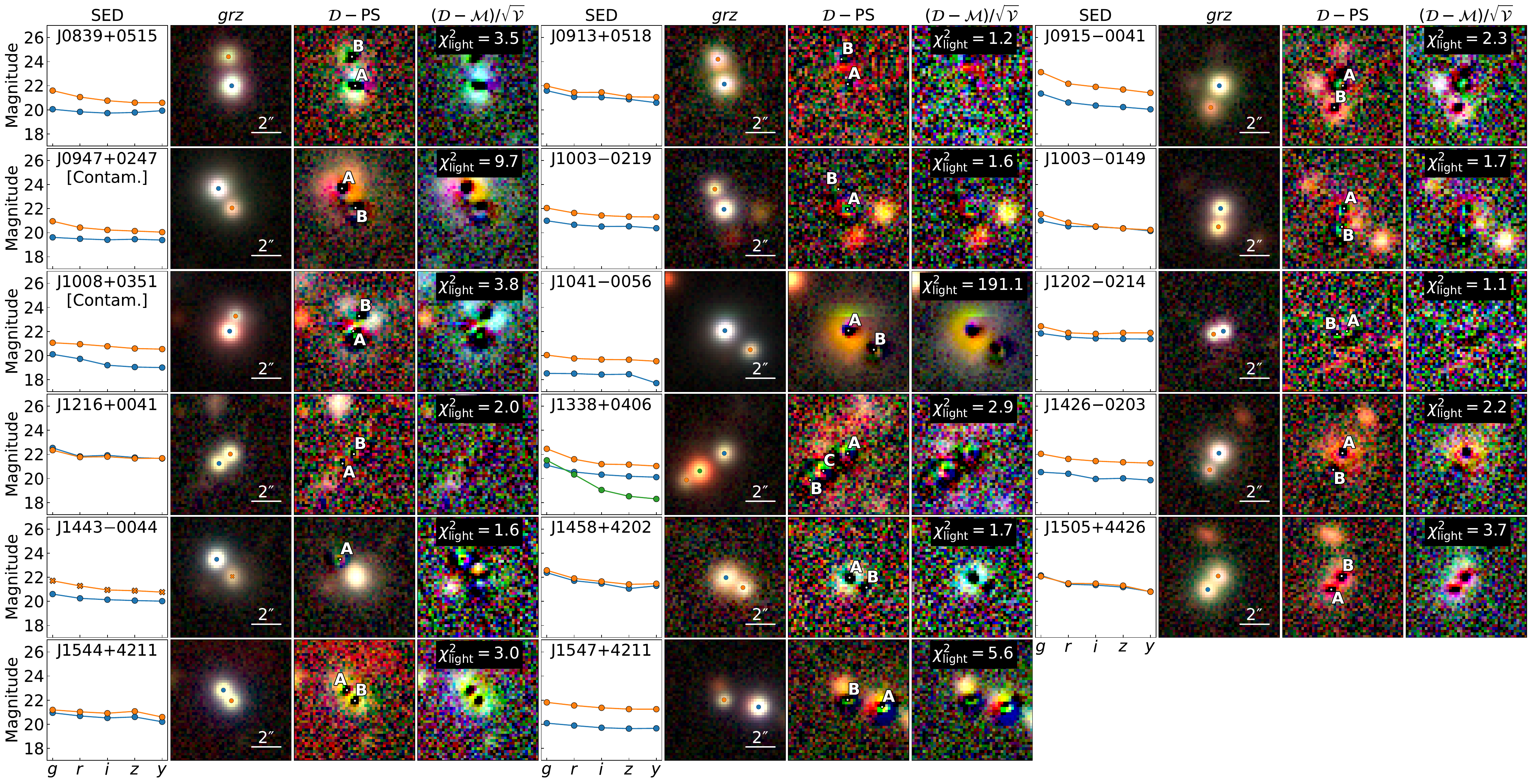}
\vskip -10pt
\caption{
Lensed candidates classified as grade B.
The labels are the same as \fref{fig:cand_A}.
}
\label{fig:cand_B}
\end{figure*}
\begin{figure*}
\centering
\includegraphics[scale=0.24]{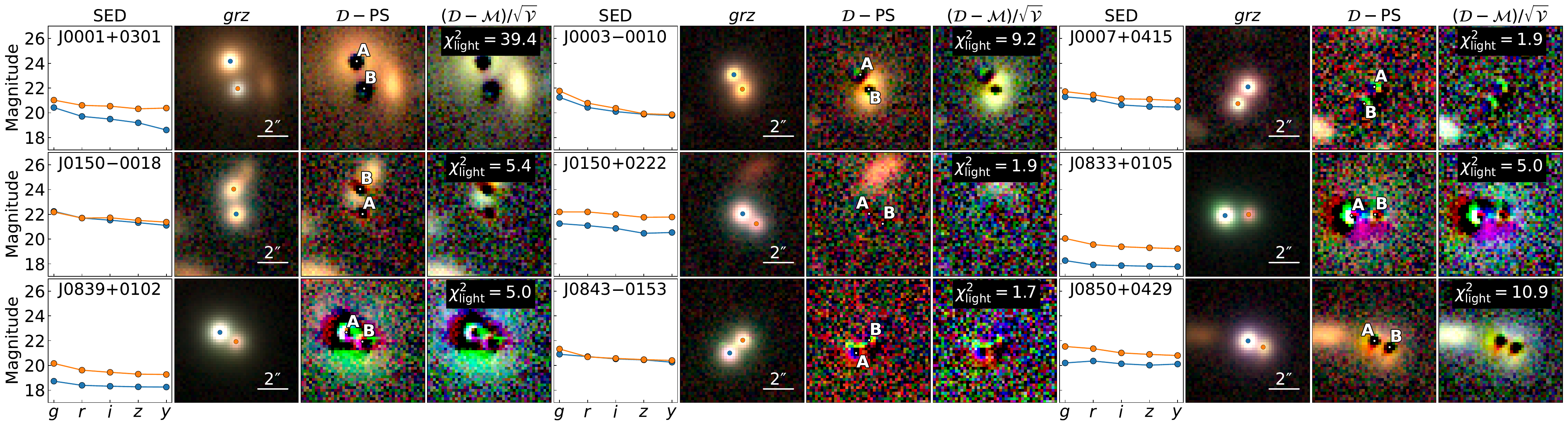}
\vskip -10pt
\caption{
Lensed candidates classified as grade C.
The labels are the same as \fref{fig:cand_A}.
}
\label{fig:cand_C}
\end{figure*}
\begin{figure*}
\ContinuedFloat
\centering
\includegraphics[scale=0.24]{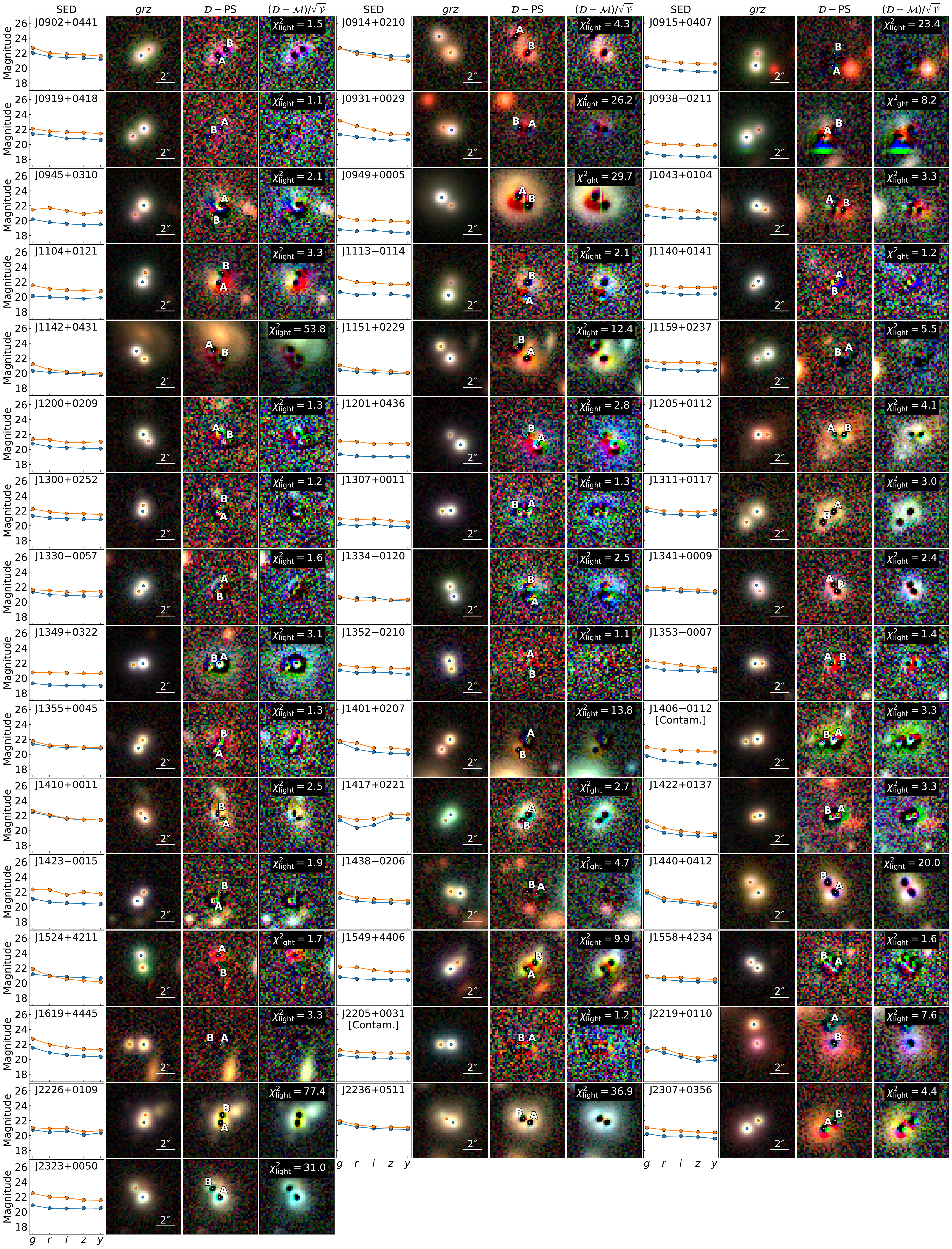}
\vskip -10pt
\caption{
\textit{Continued.}
}
\end{figure*}
\onecolumn
\begin{center}
\begin{longtable}{lrrrrcr}
\caption{Lensed candidates classified as grade A.
The preselected quasar catalogs are denoted as "MQ" for MILLIQUAS, "AW" for AllWISEAGN, and "HW" for color-cut selection using the HSC+unWISE photometry.
The average grades of the visual inspection are listed as $\Gav$.
The image separation is measured via light modeling.
The references and grades in "comment" are determined in the previous lens search works using the DECaLS data.
\textbf{The candidates as potential quads are highlighted with $\dagger$ sign following their "Name" and the confirmed contaminants are denoted in "Comment".}
}\\
\label{tab:candA}\\
\hline\hline
\multicolumn{1}{c}{Name} & \multicolumn{1}{c}{R.A [$\deg$]} & \multicolumn{1}{c}{Dec. [$\deg$]} & Catalog & $\Gav$ & sep. [$\arcsec$] & \multicolumn{1}{c}{Comment} \\
\hline
\endfirsthead
\caption{\textit{Continued.}}\\
\hline\hline
\multicolumn{1}{c}{Name} & \multicolumn{1}{c}{R.A [$\deg$]} & \multicolumn{1}{c}{Dec. [$\deg$]} & Catalog & $\Gav$ & sep. [$\arcsec$] & \multicolumn{1}{c}{Comment} \\
\hline
\endhead
\hline
\endfoot
HSC         J0015$+$0145 & $       3.786619$ & $       1.754679$ &         HW &  2.0 &    $1.010$ &                                         \\
HSC         J0025$-$0027 & $       6.253408$ & $      -0.456018$ &         HW &  3.0 &    $2.204$ &                                         \\
HSC         J0047$+$0322 & $      11.912319$ & $       3.375557$ &         HW &  3.0 &    $1.542$ &                             [3]: grade A\\
HSC         J0101$-$0130 & $      15.290820$ & $      -1.501892$ &         HW &  2.0 &    $1.927$ &                                         \\
HSC         J0117$+$0056 & $      19.481317$ & $       0.944474$ &         HW &  2.5 &    $1.988$ &                                         \\
HSC J0122$+$0037$^\dagger$ & $      20.682424$ & $       0.626798$ &         HW &  2.5 &    $1.628$ &                                         \\
HSC J0126$+$0136$^\dagger$ & $      21.688665$ & $       1.614272$ &         HW &  2.0 &    $1.939$ &                                         \\
HSC         J0130$+$0010 & $      22.562941$ & $       0.183304$ &         HW &  2.0 &    $1.021$ &                                         \\
\end{longtable}
[1]~\cite{HeEtal23}, [2]~\cite{DawesEtal22}, [3]~\cite{StorferEtal22}

\end{center}
\begin{center}
  \begin{longtable}{lrrrrcr}
\caption{Lensed candidates classified as grade B.
The notations are the same as \tref{tab:candA}.
}\\
\label{tab:candB}\\
\hline\hline
\multicolumn{1}{c}{Name} & \multicolumn{1}{c}{R.A [$\deg$]} & \multicolumn{1}{c}{Dec. [$\deg$]} & Catalog & $\Gav$ & sep. [$\arcsec$] & \multicolumn{1}{c}{Comment} \\
\hline
\endfirsthead
\caption{\textit{Continued.}}\\
\hline\hline
\multicolumn{1}{c}{Name} & \multicolumn{1}{c}{R.A [$\deg$]} & \multicolumn{1}{c}{Dec. [$\deg$]} & Catalog & $\Gav$ & sep. [$\arcsec$] & \multicolumn{1}{c}{Comment} \\
\hline
\endhead
\hline
\endfoot
HSC    J0839$+$0515 & $     129.781060$ & $       5.264559$ &         HW &  2.5 &    $1.836$ &                                         \\
HSC    J0913$+$0518 & $     138.316230$ & $       5.311164$ &         HW &  2.5 &    $1.609$ &                                         \\
HSC    J0915$-$0041 & $     138.861961$ & $      -0.693524$ &         HW &  2.5 &    $1.468$ &                                         \\
HSC    J0947$+$0247 & $     146.804584$ & $       2.795174$ & AW, MQ, HW &  2.5 &    $1.486$ &                                  Contam.\\
HSC    J1003$-$0219 & $     150.954982$ & $      -2.318416$ &         HW &  3.0 &    $1.390$ &                                         \\
HSC    J1003$-$0149 & $     150.993709$ & $      -1.831603$ &         HW &  2.5 &    $1.160$ &                                         \\
HSC    J1008$+$0351 & $     152.248168$ & $       3.851162$ & AW, MQ, HW &  2.5 &    $1.016$ &                                  Contam.\\
HSC    J1041$-$0056 & $     160.345191$ & $      -0.938461$ & AW, MQ, HW &  2.5 &    $1.997$ &                                         \\
\end{longtable}
[1]~\cite{HeEtal23}, [2]~\cite{DawesEtal22}, [3]~\cite{StorferEtal22}

\end{center}
\begin{center}
  \begin{longtable}{lrrrrcr}
\caption{Lensed candidates classified as grade C.
The notations are the same as \tref{tab:candA}.
}\\
\label{tab:candC}\\
\hline\hline
\multicolumn{1}{c}{Name} & \multicolumn{1}{c}{R.A [$\deg$]} & \multicolumn{1}{c}{Dec. [$\deg$]} & Catalog & $\Gav$ & sep. [$\arcsec$] & \multicolumn{1}{c}{Comment} \\
\hline
\endfirsthead
\caption{\textit{Continued.}}\\
\hline\hline
\multicolumn{1}{c}{Name} & \multicolumn{1}{c}{R.A [$\deg$]} & \multicolumn{1}{c}{Dec. [$\deg$]} & Catalog & $\Gav$ & sep. [$\arcsec$] & \multicolumn{1}{c}{Comment} \\
\hline
\endhead
\hline
\endfoot
HSC    J0001$+$0301 & $       0.489217$ & $       3.029465$ & AW, MQ, HW &  2.0 &    $1.730$ &                                         \\
HSC    J0003$-$0010 & $       0.923763$ & $      -0.169828$ &         HW &  2.0 &    $1.025$ &                                         \\
HSC    J0007$+$0415 & $       1.816698$ & $       4.262469$ &         HW &  2.0 &    $1.193$ &                                         \\
HSC    J0150$-$0018 & $      27.647481$ & $      -0.315818$ &         HW &  2.0 &    $1.530$ &                                         \\
HSC    J0150$+$0222 & $      27.681565$ & $       2.377538$ &         HW &  2.0 &    $1.033$ &                                         \\
HSC    J0833$+$0105 & $     128.464891$ & $       1.098478$ &         HW &  2.0 &    $1.444$ &                                         \\
HSC    J0839$+$0102 & $     129.949813$ & $       1.044056$ &         HW &  2.0 &    $1.132$ &                                         \\
HSC    J0843$-$0153 & $     130.782963$ & $      -1.886293$ &         HW &  2.0 &    $1.118$ &                                         \\
\end{longtable}
[1]~\cite{HeEtal23}, [2]~\cite{DawesEtal22}, [3]~\cite{StorferEtal22}

\end{center}

\clearpage
\begin{appendix}

\setcounter{figure}{0}
\renewcommand{\thefigure}{A\arabic{figure}}
\setcounter{table}{0}
\renewcommand{\thetable}{A\arabic{table}}
%
\begin{figure*}
\centering
\includegraphics[scale=0.24]{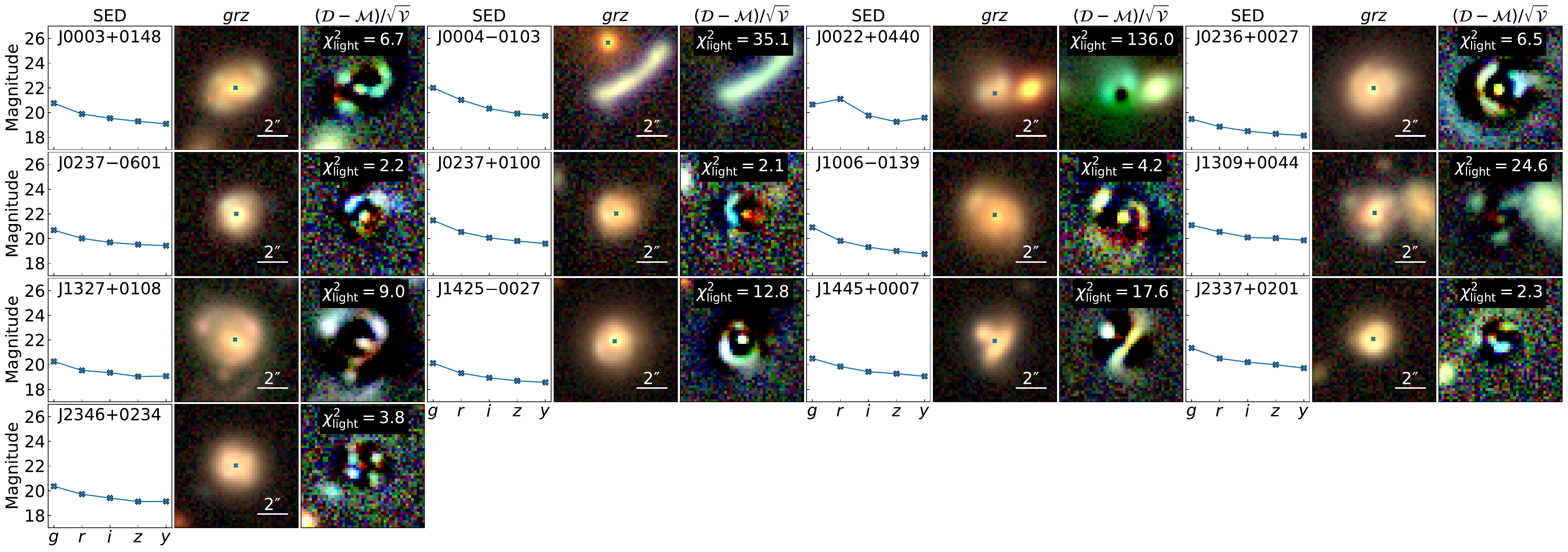}
\caption{
Lensed candidates classified as grade G.}
\label{fig:candG}
\end{figure*}
\begin{center}
  \begin{longtable}{lrrrrcr}
\caption{Lensed candidates classified as grade G.}\\
\label{tab:candG}\\
\hline\hline
\multicolumn{1}{c}{Name} & \multicolumn{1}{c}{R.A [$\deg$]} & \multicolumn{1}{c}{Dec. [$\deg$]} & Catalog & $\Gav$ & sep. [$\arcsec$] & \multicolumn{1}{c}{Comment} \\
\hline
\endfirsthead
\caption{\textit{Continued.}}\\
\hline\hline
\multicolumn{1}{c}{Name} & \multicolumn{1}{c}{R.A [$\deg$]} & \multicolumn{1}{c}{Dec. [$\deg$]} & Catalog & $\Gav$ & sep. [$\arcsec$] & \multicolumn{1}{c}{Comment} \\
\hline
\endhead
\hline
\endfoot
HSC    J0003$+$0148 & $       0.906219$ & $       1.809688$ &         HW &  2.0 &          - &                                         \\
HSC    J0004$-$0103 & $       1.215422$ & $      -1.055847$ &         HW &  3.0 &          - &                       [1]; [2]; [3]; [4]\\
HSC    J0022$+$0440 & $       5.588780$ & $       4.676942$ &         HW &  2.0 &          - &                                         \\
HSC    J0236$+$0027 & $      39.113801$ & $       0.452857$ &         HW &  2.0 &          - &                                         \\
HSC    J0237$-$0601 & $      39.306872$ & $      -6.019017$ &         HW &  2.0 &          - &                                         \\
HSC    J0237$+$0100 & $      39.351162$ & $       1.003299$ &         HW &  2.0 &          - &                                         \\
HSC    J1006$-$0139 & $     151.658371$ & $      -1.656651$ &         HW &  2.0 &          - &                                         \\
HSC    J1309$+$0044 & $     197.255413$ & $       0.735544$ &         HW &  2.0 &          - &                                         \\
\end{longtable}
[1]~\cite{RigbyEtal14}, [2]~\cite{DiehlEtal17}, [3]~\cite{JaelaniEtal20}, [4]~\cite{RojasEtal22}

\end{center}
\setcounter{table}{0}
\renewcommand{\thetable}{B\arabic{table}}
\begin{center}
  \begin{longtable}{ccrrrrrrr}
\caption{Result of light modeling for the candidates with $\Gav\geq2$.}\\
\label{tab:model}\\
\hline\hline
\multicolumn{1}{c}{Name} & \multicolumn{1}{c}{Comp.} & \multicolumn{1}{c}{$\Delta\alpha$ [$\arcsec$]} & \multicolumn{1}{c}{$\Delta\delta$ [$\arcsec$]} & HSC-$g$ & HSC-$r$ & HSC-$i$ & HSC-$z$ & HSC-$y$ \\
\hline
\endfirsthead
\caption{\textit{Continued.}}\\
\hline\hline
\multicolumn{1}{c}{Name} & \multicolumn{1}{c}{Comp.} & \multicolumn{1}{c}{$\Delta\alpha$ [$\arcsec$]} & \multicolumn{1}{c}{$\Delta\delta$ [$\arcsec$]} & HSC-$g$ & HSC-$r$ & HSC-$i$ & HSC-$z$ & HSC-$y$ \\
\hline
\endhead
\hline
\endfoot
HSC     J0015$+$0145 [A] & $  A$ & $    -0.056$ & $    -0.365$ &$     20.45$ & $     20.07$ & $     20.03$ & $     20.00$ & $     19.82$\\
                         & $  B$ & $     0.115$ & $     0.631$ &$     21.56$ & $     21.04$ & $     20.98$ & $     20.71$ & $     20.44$\\
                         & $  G$ & $    -0.191$ & $     0.331$ &$     20.74$ & $     19.76$ & $     19.33$ & $     19.10$ & $     18.78$\\
HSC     J0025$-$0027 [A] & $  A$ & $     0.286$ & $    -1.231$ &$     23.98$ & $     23.06$ & $     22.47$ & $     21.98$ & $     21.49$\\
                         & $  B$ & $     0.627$ & $     0.947$ &$     25.65$ & $     23.52$ & $     22.34$ & $     21.86$ & $     21.46$\\
                         & $  G$ & $    -0.053$ & $    -0.095$ &$     20.79$ & $     20.29$ & $     19.95$ & $     19.76$ & $     19.67$\\
HSC     J0047$+$0322 [A] & $  A$ & $     0.079$ & $    -0.275$ &$     21.86$ & $     21.79$ & $     21.61$ & $     21.38$ & $     21.41$\\
                         & $  B$ & $    -0.785$ & $     1.003$ &$     22.06$ & $     22.01$ & $     21.89$ & $     21.63$ & $     21.68$\\
\end{longtable}

\end{center}

\end{appendix}


\bsp	
\label{lastpage}
\end{document}